\documentclass{aa}  

\usepackage{graphicx}
\usepackage{txfonts}
\usepackage{gensymb}
\usepackage{xcolor}
\begin{document} 
   \title{The complex case of MACS~J0717.5+3745 and its extended filament: intra-cluster light, galaxy luminosity function, and galaxy orientations}

   \titlerunning{MACS~J0717.5+3745 and its extended filament}

   \author{A. Ellien
          \inst{1}
          \and
          F. Durret
          \inst{1}
          \and
          C. Adami
          \inst{2}
          \and
          N. Martinet
          \inst{2}
          \and
          C. Lobo
          \inst{3,4}
          \and
          M. Jauzac
          \inst{5,6,7}
          }

   \institute{Sorbonne Université, CNRS, UMR 7095, Institut d'Astrophysique de Paris, 98bis Bd Arago, F-75014 Paris
            \and
             Aix-Marseille Univ., CNRS, CNES, LAM, Marseille, France
            \and
            Instituto de Astrof\'{\i}sica e Ci\^encias do Espa\c co, Universidade do Porto, CAUP, Rua das Estrelas, PT4150-762 Porto, Portugal
            \and
        Departamento de F\'{\i}sica e Astronomia, Faculdade de Ci\^encias, Universidade do Porto, Rua do Campo Alegre 687, PT4169-007 Porto, Portugal
        \and
        Centre for Extragalactic Astronomy, Department of Physics, Durham University, Durham DH1 3LE, United Kingdom
        \and
        Institute for Computational Cosmology, Durham University, South Road, Durham DH1 3LE, UK
        \and
        Astrophysics and Cosmology Research Unit, School of Mathematical Sciences, University of KwaZulu-Natal, Durban 4041, South Africa
             }

   \date{Received April 12, 2019; accepted May 23, 2019}

 
   \abstract
   {The properties of galaxies are known to be affected by their environment, but although galaxies in clusters and groups have been quite thoroughly investigated, little is known presently on galaxies belonging to filaments of the cosmic web, and on the properties of the filaments themselves.}
   {We investigate here the properties of the rich cluster MACS~J0717.5+3745  and its extended filament, by analyzing the distribution and fractions of intra-cluster light (ICL) in the core of this cluster and by trying to detect intra-filament light (IFL) in the filament. We analyze the galaxy luminosity function (GLF) of the cluster core and of the filament. We also study the orientations of galaxies in the filament to better constrain the filament properties.}
   {This work is based on \emph{Hubble Space Telescope} archive data, both from the Hubble Frontier Fields in the F435W, F606W, F814W, and F105W bands, and from a mosaic of images in the F606W and F814W bands. The spatial distribution of the ICL is determined with our new wavelet-based software, \texttt{DAWIS}. The GLFs are extracted in the F606W and F814W bands, with a statistical subtraction of the background, and fit with Schechter functions. The galaxy orientations in the filaments are estimated with SExtractor after correction for the Point Spread Function.}
   {We detect a large amount of ICL in the cluster core, but no IFL in the cosmic filament. The fraction of ICL in the core peaks in the F606W filter before decreasing with wavelength. Though quite noisy, the GLFs in the filament are notably different from those of field galaxies, with a flatter faint end slope and an excess of bright galaxies. We do not detect a significant alignment of the galaxies in the filament region that was analyzed.}
   {}


   \maketitle

%

\section{Introduction}
\label{sec:intro}

Already in the early 1980's, \citet{Zeldovich1982} have predicted through theoretical models of structure formation that small fluctuations from the early universe would lead to a distribution of matter condensed along filaments, sheets and voids. This resulted in the cosmic web that is now the framework for cosmology, as e.g. described by \citet{Bond1996}. 
The detection of filaments is also interesting since, as underlined by \citet{Eckert2015} they could account for the missing baryons in the universe.
However, it remains difficult to detect cosmic web filaments in real data, as summarized in \citet{Libeskind2018}, who compare twelve different methods to identify and classify the cosmic web. Weak lensing has been a way to detect filaments between clusters, as shown by \citet{Dietrich2012}, in particular between clusters forming a pair \citep{Planck2013}. The orientations of red galaxies have also been used as probes of filaments by \citet{Rong2016}.  \citet{Stoica2005} and \citet{Tempel2016} built the \textsc{Bisous} filament finder, a marked point process built to model multi-dimensional patterns, now publicly available. 
Another recent approach is the search for large scale diffuse radio emission, as recently detected by \citet{Vacca2018}, who believe this emission is linked to a large-scale filament of the cosmic web likely associated with an over-density traced by nine massive clusters. 

The properties of galaxies in filaments have only started to be investigated a few years ago. Based on the large-scale HORIZON-AGN hydrodynamical cosmological simulation, \citet{Dubois2014} have found that at $1.2<z<1.8$ low mass blue star-forming galaxies have a spin preferentially aligned with their neighbouring filaments, while high mass red quiescent galaxies tend to have a spin perpendicular to nearby filaments. They interpret the reorientation of galaxies as due to galaxy mergers, and find that the mass transition occurs around $3\times 10^{10}$~M$_\odot$. Based on simulations, comparable results were found by  \citet{Ganesh2018}, who also found that the transition mass between the two regimes increases with increasing filament diameter. \citet{Wang2018}, instead, found that the transition mass decreases with increasing redshift.

As far as observational data are concerned, the application of \textsc{Bisous} to SDSS data also led to several interesting results, among which we emphasize three. First, \citet{Tempel2013a} found that the
minor axes of ellipticals are preferentially perpendicular to their hosting filaments. Second, \citet{Tempel2014} published a
public catalogue of filaments detected with \textsc{Bisous}, the longest filaments they detect reaching 60~$h^{-1}$~Mpc.  Third, \citet{Tempel2015} discovered a statistically significant alignment between the satellite galaxy position and the filament axis, the alignment being stronger for the reddest and brightest central and satellite galaxies.

Based on the SDSS data release 12, \citet{Chen2017} found  results consistent with those based on numerical simulations: red or high mass galaxies tend to reside closer to filaments than blue or low mass galaxies. The star formation rate of galaxies in the outskirts of clusters was also found to be higher \citep{Mahajan2012}.
\citet{Kuutma2017} investigated the impact of filament environment on galaxies, quantifying the environment as the distance to the spine of the nearest filament. They find an increase of the elliptical to spiral ratio while moving from voids to filament spines, but they do not detect an increase in the galaxy stellar mass while approaching filaments. They interpret their results as due to an increase in the galaxy-galaxy merger rate and/or to the cutoff of gas supplies near and inside the filaments. This study suggests that cosmic web filaments must have an impact on galaxy properties. Indeed,  \citet{Sarron2019} have searched for filaments around the clusters detected by the AMASCFI software in the Canada France Hawaii Telescope Legacy Survey and showed, among other results, that pre-processing in filaments  could occur in galaxy groups located in the filaments.

Galaxy luminosity functions (GLFs) can be an interesting tool to understand better the properties of galaxies in different environments. GLFs have been analysed individually for quite a large number of clusters in a wide range of redshifts these last decades \citep{Smail1998,DeLucia2004,Andreon2006,DeLucia2007,Rudnick2009,Vulcani2011,DePropris2013,Martinet2015,Zenteno2016,Martinet2017}. 
They have also been studied in a few very large samples of clusters, that allow to stack GLFs, and to analyse their variations with cluster mass or redshift, separating red and blue galaxies, as recently done for example by  \citet{Ricci2018} or \citet{Sarron2018}.
To our knowledge, GLFs have not yet been estimated in cosmic filaments, but they are expected to have properties intermediate between GLFs in dense environments such as groups or clusters, and GLFs of field galaxies. We will discuss here the properties of the GLFs in the zones covered by the filament.

In parallel, the build-up of Intra-Cluster Light (ICL) in the current hierarchical model of evolution of big structures in the Universe is another field of interest when looking at cosmic filaments. First mentioned by \citet{Zwicky1951} as he discovered an extended low surface brightness luminous halo around the Brightest Cluster Galaxy (BCG) in the Coma cluster, the ICL has been a growing field of research through recent years. Numerous studies have been made to investigate the nature and properties of this diffuse optical component, and it is now commonly admitted that the ICL is composed of stars that are not gravitationally bound to any cluster galaxy and are more related to the global gravitational potential of the cluster. 

ICL has been found in nearby galaxy clusters such as the Virgo cluster \citep{Mihos2017} or the Coma cluster \citep{Gu2018,Jimenez-Teja2019}, in the form of an extended luminous halo centered on the BCG, superposed on a variety of substructures such as straight streams \citep{Mihos2005}, curved arcs \citep{Trentham1998}, large plumes \citep{Gregg1998} or tidal tails \citep{Krick2006,Janowiecki2010}. However, due to the difficulty to quantify their morphological properties, the studies of such features remain mainly qualitative, as a disclosure of the numerous mechanisms occurring in the ICL. At intermediate redshifts ($0.1<z<1$), large samples of galaxy clusters are available, allowing systematic approaches to quantify the physical properties of the smooth ICL halo  \citep{Krick2007,Guennou2012,DeMaio2018,Jimenez-Teja2018,Montes2018,Zhang2019}. These observational studies have allowed us to accumulate knowledge on the ICL under various forms, such as the fraction of ICL in galaxy clusters (from 10 to 50\%), its color, velocity, metallicity, and spatial distribution. Observations of ICL have also been made in high-redshift ($z>1$) young galaxy clusters \citep{Adami2013,Ko2018}, indicating a correlation between the dynamical evolution of galaxy clusters at early-times and the evolution of their ICL.

While the presence of ICL in galaxy clusters is not questioned any more, its formation mechanisms are still under discussion.  
Based on numerical simulations, \citet{Merritt1984} showed that the ICL could form from stars dynamically stripped from their parent galaxy. 
Two main processes have been proposed:  
tidal stripping by the galaxy cluster gravitational potential \citep{Byrd1990}, or violent encounters between a pair or a group of galaxies, 
leading to the formation of large tidal streams which then mix into the 
ICL component \citep{Moore1996,Moore1999,Mihos2004a}. Numerical works based on $N$-body, and hydrodynamical cosmological simulations or semi-analytic model simulations have investigated directly the effect of those formation mechanisms \citep{Napolitano2003,Willman2004,Murante2004,Murante2007,Rudick2006,Rudick2009,Contini2014,Contini2018}. While the literature is overall consistent with the fact that a large fraction of stars is found in the ICL at $z=0$, and that the mergers and violent interactions between galaxies seem to be the main providers of those stars, there are still great discrepancies among the results, such as the time-period in which the ICL forms, its formation rate, or the relation between cluster mass and ICL fraction \citep{Rudick2011,Tang2018}.

Galaxy clusters are not the only place where diffuse light material is created. \citet{Sommer-larsen2006} showed in hydrodynamical simulations that galaxy groups act like smaller scale galaxy clusters, producing their own Intra-Group Light (IGL) through merging processes. This has been confirmed by several observational studies that found evidence for IGL in groups of galaxies \citep{DaRocha2005,Aguerri2006,DaRocha2008}. One could imagine that, since cosmic filaments also seem to feature a significant amount of galaxy-galaxy mergers \citep{Kuutma2017}, an extended and smooth component, the Intra-Filament Light (IFL) could be produced in the same manner.

A way of increasing the ICL of a galaxy cluster could be through mergers with galaxy groups. This process has also been proposed by \citet{Mihos2004b} as a potential source of ICL, as IGL could be formed in groups falling into bigger structures. In their work on the formation of ICL through tidal streams, \citet{Rudick2009} also showed that dynamically cold tidal streams could be formed through violent galaxy encounters in groups at early times. In some cases the tidal potential of such groups would be too weak for the streams to relax into the smooth IGL component, and the streams would stay still until their galaxy group falls into a stronger gravitational potential, to be finally mixed in the associated ICL. In the current hierarchical model of large scale structure evolution, such groups featuring large and bright tidal streams that could become ICL in the future should be found in cosmic filaments during their infall into galaxy clusters. However, such systems have not been identified yet, due to the difficulty to detect and characterize both cosmic filaments and ICL.
 
MACS~J0717.5+3745 (hereafter MACS~J0717) is a cluster at a redshift z=0.5458, known to be the most massive cluster at $z>0.5$, with a mass  M$_{200}=23.6\times 10^{14}\ h_{70}^{-1}$~M$_\odot$ \citep{Martinet2016}. 
Based on HFF data, \citet{Diego2015} and \citet{Limousin2016} made a mass reconstruction of the cluster.
MACS~J0717 is embedded in a very long double filament of galaxies extending over more than 9~Mpc in total \citep{Ebeling2004,Kartaltepe2008,Jauzac2012,Durret2016}. As already noted by \citet{Durret2016}, 
neither of the two filaments (that they labeled B and C) is strongly detected in the X-rays, suggesting that we are probably dealing with cosmic web filaments linked to the cluster rather than with clusters or groups merging at large scales. This led us to choose this system to study for the first time the properties of a large scale filament that seems to be feeding a massive cluster: the spatial distribution of the ICL, the galaxy luminosity function (GLF), which will be compared with that of the cluster itself \citep{Martinet2017} and to field galaxy GLFs, and the orientations of the filament galaxies. The fact that MACS~J0717 is covered by the Hubble Frontier Fields \citep{Lotz2017} and therefore very deeply observed with the \emph{HST} makes it an ideal object to attempt the detection of diffuse intracluster light (ICL) with our new software, \texttt{DAWIS} (Detection Algorithm with Wavelets for Intra-cluster light Surveys).

The paper is organized as follows. In Sect.~\ref{sec:ICL} we describe our analysis of the ICL. In Sect.~\ref{sec:GLF} we present the galaxy luminosity function in the filament. In Sect.~\ref{sec:PAs} we analyse the orientations of the galaxies in the filament. All these results are discussed in Sect.~\ref{sec:conclusions}.


\section{The data}
\subsection{The Hubble Frontier Fields}
\label{HFF}

MACS~J0717 is part of the Hubble Frontier Fields Survey (HFF)\footnote{https://frontierfields.org/meet-the-frontier-fields/macsj0717/}, the deepest \emph{Hubble} broadband photometric survey dedicated to galaxy clusters today (ID13498, PI: J.Lotz). Long exposure images of six massive galaxy clusters in the redshift range $0.3<z<0.55$ and their parallel fields were taken in a two-step process. Epoch1 of the observing campaign for MACS~J0717 took place between 2014 September and 2014 December, while Epoch2 took place between 2015 February and 2015 March \citep{Lotz2017}. 
The optical data were obtained with the Advanced Camera for Surveys (ACS). The IR channel of the Wide Field Camera (WFC3/IR) was used to obtain NIR images in four filters (F105W, F125W, F140W, F160W). 
Since the main goal of the paper is the ICL detection, we strategically chose four \emph{HST} bands: F435W (UV rest frame, sensitive to star formation at the cluster redshift), F606W (including the [OII]$\lambda$3727 line at the cluster redshift), F814W (the most sensitive ACS band and including the [OIII emission lines] at the cluster redshift), and F105W (including the H$\alpha$ and [SII] lines at the cluster redshift). Details on the data reduction can be found in the archive handbook\footnote{https://archive.stsci.edu/pub/hlsp/frontier/}.

Here we retrieve the F105W, F435W, F606W and F814W images of the core of MACS~J0717, and its parallel field from the public archive\footnote{https://archive.stsci.edu/pub/hlsp/frontier/macs0717/images/hst/}. Two pixel sizes are available, 0.03\arcsec\ and 0.06\arcsec. We choose the 0.06\arcsec\ pixel size to increase the sensitivity and detect low surface brightness objects.
The F606W and F814W filters are chosen to match the \emph{HST} mosaic filters covering the filament of MACS~J0717 (see Section~\ref{HSTmosaic}), and the F435W and F105W filters are also retrieved to explore the behaviour of the ICL in respectively bluer and redder filters. Those images are used to detect ICL in the core of the MACS~J0717 (see Section~\ref{ICLcore}).

\begin{figure}
  \centering
  \includegraphics[width=3.5in,clip=true]{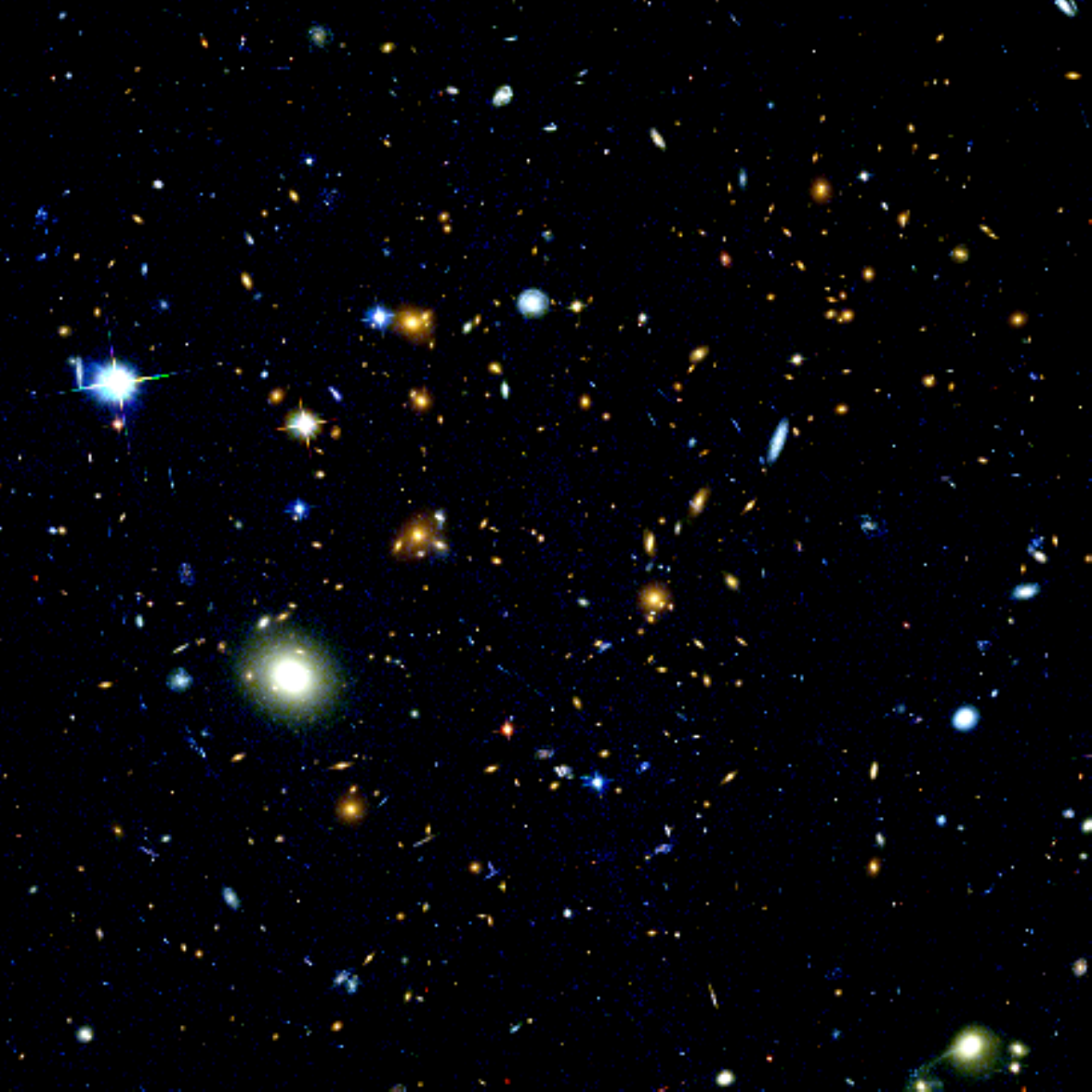}
  \caption{RGB image of the core of MACS~J0717 in the Hubble Frontier Field (F435W, F606W, F814W). The size of the image is 2.8$\times$2.8 arcmin$^2$ (corresponding to $\sim 1\times 1$ Mpc$^2$ at the cluster redshift).}
  \label{fig:HFF}
\end{figure}

\subsection{HST Mosaic}
\label{HSTmosaic}

\begin{figure}
  \centering
  \includegraphics[width=3.5in,clip=true]{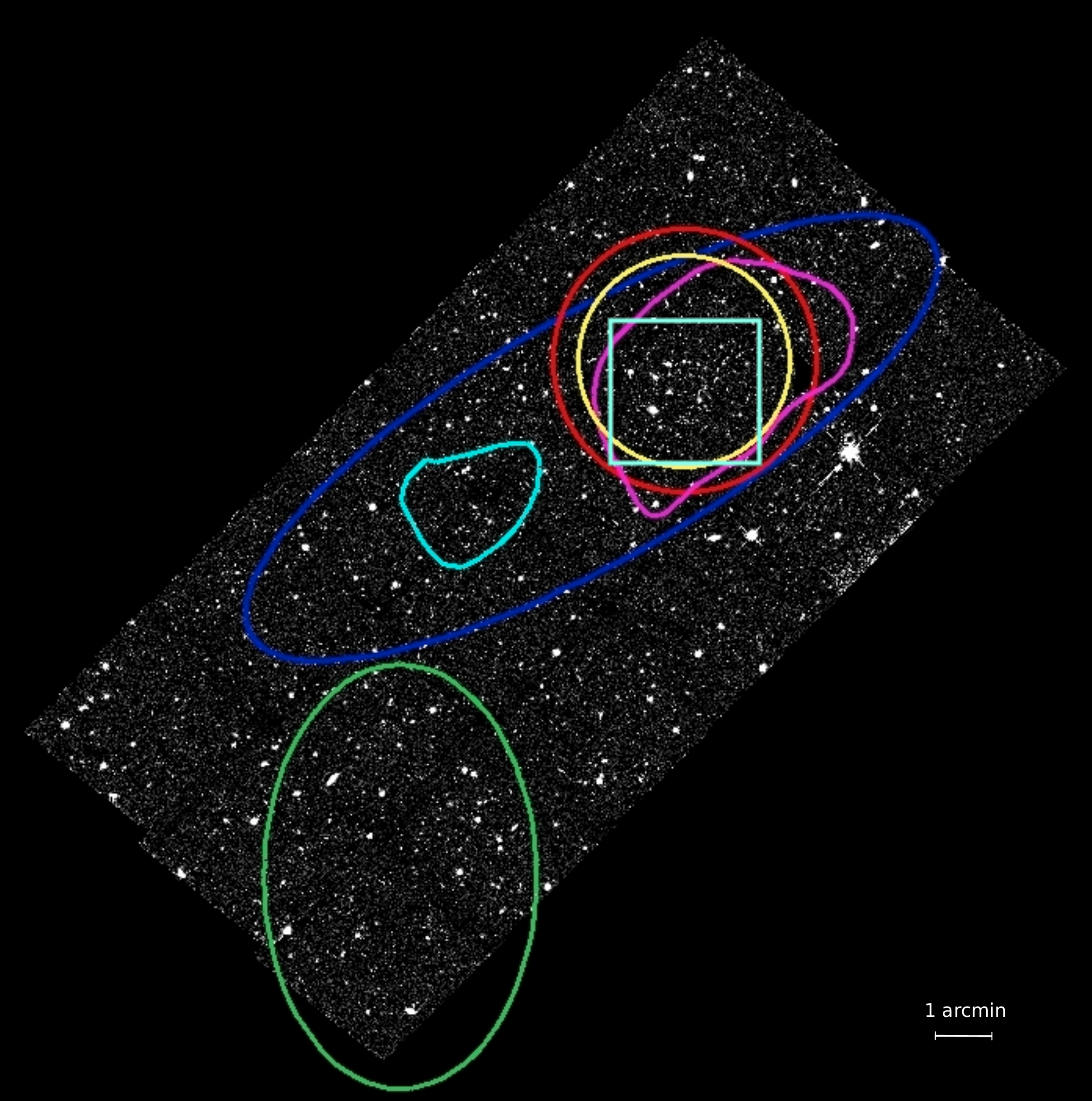}
  \caption{Full HST mosaic, covering the entire field of MACS~J0717 and its extended filament. The blue ellipse corresponds to region B, the cosmic filament area closest to the cluster. The green ellipse corresponds to region C, the cosmic filament area south-east of the cluster and located further. The red and yellow circles are respectively 1~Mpc and 807~kpc radius circles centered on the core of MACSJ~0717. The cyan and magenta contours show 3$\sigma$ weak lensing contours from \citet{Martinet2016}. The cyan rectangle is the HFF field-of-view in the optical filters - see Fig~\ref{fig:HFF} for a zoom on this area.}
  \label{fig:HSTmosaic}
\end{figure}

The HFF (see Fig.~\ref{fig:HFF}) does not cover the full cosmic filament to the south-east of MACS~J0717 (see Fig.~\ref{fig:HSTmosaic}). We therefore use another set of \emph{HST} images obtained between 2005 January 9, and February 9 with the ACS (GO-10420, PI: Ebeling). This mosaic consists in 18 images of pixel size 0.05\arcsec\ in the F606W and F814W filters, and has been used in past works to detect the cosmic filament. More details can be found in \citet{Jauzac2012}. This mosaic is used to compute the Galaxy Luminosity Functions (GLF) (see Section~\ref{sec:GLF}), to study the orientations of galaxies (see Section~\ref{sec:PAs}) and to look for IFL and tidal streams in the filament (see Section~\ref{sec:tidalstreams}). A global view of the mosaic can be seen in Fig.~\ref{fig:HSTmosaic}.

\section{Intra-cluster light}
\label{sec:ICL}

We describe here our study of intracluster light in MACS~J0717 and its filament. We first present our software in Sect.~\ref{DAWIS}. We then apply our method to the HFF data to detect intra-cluster light (ICL) in the core of MACSJ~0717 in Sect.~\ref{ICLcore}. Finally, we describe our search for intra-filament light (IFL) under various forms in the filament of MACS~J0717 in Sect.~\ref{sec:tidalstreams}.

\subsection{Presenting \texttt{DAWIS}}
\label{DAWIS}

The ICL is a very extended and diffuse light source. Numerous instrumental factors can impact its detection, such as scattered light in telescopes, flat-fielding uncertainties or background level estimation. The procedure to constrain these elements is given in Section~\ref{ICLcore}. Other astronomical effects are also difficult to take into account, and separation of ICL from galaxy light is always a challenge. In the past decades, different methods have been applied to perform this kind of analysis: fitting and extraction of galaxy emission using light profiles \citep{Vilchez-gomez1994, Gonzales2005, Jimenez-Teja2016, Jimenez-Teja2018, Jimenez-Teja2019}, raw masking of sources with pixel values greater than an estimated detection threshold \citep{Burke2012, DeMaio2018, Ko2018, Montes2018}, or using wavelet packages to model and remove galaxy light components \citep{Adami2005, DaRocha2005,  DaRocha2008, Guennou2012, Adami2013}. 

While every approach has its advantages and disadvantages, the wavelet one is particularly flexible and efficient to disentangle bright sources from low surface brightness diffuse ones, making it exceptionally well adapted to the detection of ICL. Indeed, contrary to fitting-oriented methods, a wavelet analysis does not need any prior information to perform detection and modeling of objects, but on the other hand its adaptability requires extended CPU-time computing. With this in mind, we create \texttt{DAWIS} (Detection Algorithm with Wavelets for Intra-cluster light Surveys), a highly parallelized wavelet-based detection package created specially for the detection and study of ICL. \texttt{DAWIS} is optimized to run on large images faster than regular linear wavelet packages, in order to process large amounts of data.

In the following sub-sections, we give a global description of \texttt{DAWIS}, which consists in the wavelet convolution of an astronomical image, the detection of objects in wavelet space, and the reconstruction of these  detected objects. More detailed explanations and tests on simulations can be found in Ellien et al. (in preparation).

\subsubsection{Wavelet convolution}

Astronomical images can be hierarchically decomposed: bright compact sources such as stars or galaxy cores are located in larger envelopes like star halos or galaxy disks, which are themselves enclosed in very large, diffuse low surface brightness sources such as the ICL. All these objects are projected on the sky background, the largest component (covering the entire image) setting the surface brightness detection limit. A multi-scale approach is ideal to disentangle these different luminosity scales and to study them. 

\texttt{DAWIS} is based on Mallat's \textit{\`{a} trous} wavelet algorithm \citep{Shensa1992}, which is particularly suited for astronomical images. It is a fast discrete redundant wavelet transform respecting flux conservation. Going through this algorithm, an image is convolved in $N_{\rm{lvl}}$ wavelet planes, following a multi-scale vision as described in \citet{Bijaoui1995}. The process is done by smoothing iteratively $N_{\rm{lvl}}+1$ times the original image with a varying B-spline kernel, the difference between two successive smoothed images giving a wavelet plane. Each wavelet plane contains features with a characteristic size of $2^{n}$ pixels, $n$ being the index of the plane ($n=0,1,2,3,...,N_{\rm{lvl}}$). Small values of $n$ correspond to bright and compact luminous features, while greater values correspond to large and low intensity ones, $n=N_{\rm{lvl}}$ corresponding to the large sky background variations. The maximal number of wavelet planes $N_{\rm{lvl,max}}$ you can get for an image is given by its size which is $2^{N_{\rm{lvl,max}}}$ pixels.

\subsubsection{Detection of objects}
\label{Detection}

After a wavelet convolution, astronomical objects are decomposed in several features through the different wavelet planes. In each plane, we apply a thresholding to determine what are the statistically significant pixels composing these features. Noise representation in wavelet space is an important attribute of this approach. Indeed, the result of a wavelet convolution of Gaussian noise is also Gaussian noise, but with a shift in intensity depending on $n$. Astronomical noise is strongly dominated by small characteristic size variations, and the first wavelet plane noise ($n=0$) is basically high intensity pixel-to-pixel noise. The noise intensity falls drastically with increasing $n$, revealing astronomical objects. This means that in large $n$ wavelet planes, extended and low surface brightness features are easily identified because of greatly reduced background noise.

The detection threshold is estimated in each wavelet plane separately: a standard deviation of the intensity, $\sigma_{n}$, is computed using a 3$\sigma$ clipping algorithm, and the significant pixels are set to have intensity values higher than $k\sigma_{n}$. This threshold is different for each wavelet plane, but the same value $k$ is applied everywhere, and we will refer from now on to these different thresholds as $k\sigma_{\rm{w}}$, the detection threshold in the wavelet space (usually $3\sigma_{\rm{w}}$ or $5\sigma_{\rm{w}}$).

After thresholding, the significant pixels are grouped in regions using scale-by-scale segmentation. We then create inter-scale trees by linking together significant regions from different planes by looking at their spatial distribution. Trees with connected regions from at least three different planes are recognized as valid representations of astronomical objects in the wavelet space.

\subsubsection{Reconstruction of objects}

For each detected inter-scale tree, the region with the highest pixel value is set as the main region. The information from this region, and every region of the tree that belongs to smaller $n$ planes, is used to reconstruct the object in real space, using a conjugate gradient algorithm \citep{Starck1998}. All the reconstructed objects are then inserted in a single image that we call the `reconstructed image'. A residual image is produced by subtracting the reconstructed image from the original astronomical one. We refer from now on to the full process of wavelet convolution, detection in wavelet space, reconstruction of every detected object, and computation of reconstructed and residual images as a `run'.

The fact that only information from the main region, and regions with smaller indexes, are used to reconstruct an object means that \texttt{DAWIS} detects and reconstructs in priority bright and compact sources. Many low surface brightness sources can be missed in this way and are found in the residual image. In that case, a second \texttt{DAWIS} run can be applied to this residual image. Objects that have not been detected in the first run are then detected, reconstructed, and subtracted from the first residual image, producing a second residual image. This iterative process can be generalized to $N$ runs if needed. For example, the number of runs can be pushed to 8 to create masks (see Section~\ref{masks}).

\subsection{ICL in the core of MACS~J0717}
\label{ICLcore}

Here we describe the data processing applied to the HFF images in order to quantify the contribution of ICL in the core of MACS~J0717 to the total luminosity budget.

\subsubsection{Background estimation}
\label{sec:background}

An accurate estimation of the sky background is essential to the study of low surface brightness features, as the subtraction of the non-uniform background could erase significant signal. The task is particularly complicated if those features are extended over large areas, since a classical global estimation of the sky background is contaminated by the light of diffuse sources. In the case of the HFF images, the Field-of-View (FoV) is smaller than the actual size of MACS~J0717 and the ICL might cover a large portion of it (potentially the full image). More sophisticated methods of background estimation are then necessary \citep{DeMaio2018,Ko2018,Montes2018}.

For the HFF data, we take advantage of the deep parallel fields available \citep{Lotz2017}, and use them to compute the sky background. The HFF parallel fields are pointed at a single target distant by $\sim$6 arcmin from the galaxy cluster core (equivalent to $\sim$2.2 Mpc at the redshift of MACS~J0717), in a region covering the galaxy field. This is far enough from the cluster centre to avoid large ICL contribution, and should provide a fair estimation of the sky background.

In the HFF images, the subtraction of a constant sky background was performed during the data reduction. In order to get rid of negative pixel values which DAWIS cannot deal with, we add a constant to every image we process during this work, and remove it from the output images at the end of the wavelet processing. For each band we create $\sim$50 random circular regions of radius 3.6\arcsec, covering the entire parallel field. This size is chosen to be larger than typical field galaxies, but small enough to avoid the large scale background variations. For each region, we use a 3$\sigma$ clipping algorithm to estimate a standard deviation of the pixel values, which allows us to remove the bright sources that can be found above the average intensity level in some of them. Then we once again apply a 3$\sigma$ clipping algorithm, but this time on the values of the standard deviation for all regions, removing outliers. The final value obtained is the global standard deviation $\sigma_{\rm{bkg}}$. We insist here on the fact that $\sigma_{\rm{bkg}}$ is different from $\sigma_{\rm{w}}$. The latter is computed in the wavelet space and used to detect objects before they are reconstructed (See Section~\ref{Detection}), while $\sigma_{\rm{bkg}}$ is used to compute the detection threshold in the final residual map after all of the wavelet processes are done. The 3$\sigma_{\rm{bkg}}$ and 5$\sigma_{\rm{bkg}}$ detection thresholds for each filter are given in Table~\ref{tab:detectionlimit}.

\begin{table*}
  \centering
  \caption{Detection threshold and ICL fractions computed for each of the four filters of the HFF. The thresholds are used to create the ICL maps from the residual images (see Fig,~\ref{mapICL}). The fractions are computed from the ICL maps and the reconstructed images in the four filters (see Section~\ref{sec:resultsICL}) within the radius that is indicated for each filter. 
 Error bars correspond to the 95\% confidence interval computed from bootstrap resampling (see text for details). The radii are the same as in \citet{Jimenez-Teja2019} for comparison purposes.}
  \begin{tabular}{l|cccc}
    \hline
    \hline
    HFF & F435W & F606W & F814W & F105W \\
   \hline
    3$\sigma_{\rm{bkg}}$ (mag.arcsec$^{-2}$) & 29.89 & 29.96 & 30.03 & 29.97 \\
    5$\sigma_{\rm{bkg}}$ (mag.arcsec$^{-2}$) & 29.34 & 29.41 & 29.50 & 29.41 \\
    Radius (kpc) & 275.3 & 562.5 & 421.5 & FoV \\
    $f_{\rm{ICL}} (\%)$ & 2.48$^{+0.19}_{-0.20}$ & 24.43$^{+3.37}_{-1.71}$ & 16.10$^{+1.03}_{-1.03}$ & 13.22$^{+1.76}_{-1.49}$ \\
    \hline
  \end{tabular}
  \label{tab:detectionlimit}
\end{table*}

\subsubsection{Point Spread Function}

The Point Spread Function (PSF) is the generic term used to characterize the response of an instrument, telescope or detector, to a point-like source, and is the combination of a variety of effects such as the diffraction pattern induced by the telescope aperture, optical aberrations, scattered light within the instrument, and atmospheric turbulence, resulting in a blurring of the source image. While we observe this deformation, the variety of these effects and their different origins make it difficult to estimate or model them accurately, and each instrument needs specific study and characterization of its PSF. 

We measure the PSF on bright non-saturated stars. While the inner part is by far the most luminous one, the wings are also important. They can be a source of contamination for low surface brightness features \citep{Sandin2014,Sandin2015}, bringing light from the inner parts of galaxies to their outer halo and modifying their colour properties and luminosity profile at large radius \citep{Capaccioli1983,deJong2008,Trujillo2016}. This effect impacts the ICL too, since light from galaxies can be artificially brought to the inter-galactic medium, simulating flux emitted by diffuse low surface brightness sources and polluting real ICL light.

Correcting an astronomical image for PSF effects is not an easy task, as PSF properties depend on e.g. wavelength, spatial position on the image, and exposure time. Consequently, the PSF of a science image is usually measured empirically, with tools like \texttt{PSFEx} \citep{Bertin2011,Bertin2013}. This method estimates the PSF directly from stars in the science image by stacking them, and can measure the PSF accurately up to radial distances of a few arcseconds. Other stacking methods to increase the signal-to-noise of the PSF wings have been used in some works \citep{Janowiecki2010,Karabal2017,DeMaio2018}, but require numerous stars and a specific observation strategy to avoid saturation.

In our case, the small field of view of the HFF images strongly reduces the number of stars available for PSF measurements. For this reason, we choose to use \texttt{TinyTim}, a modelling tool that has been created to provide PSFs for all the instruments and observation modes of the \emph{HST} \citep{Krist2011}. \texttt{TinyTim} takes into account many factors for the making of the PSF, such as aberrations, time-dependent focus or geometric distortions (more details are available in the \texttt{TinyTim} User's Guide available on the project website\footnote{http://www.stsci.edu/hst/observatory/focus/TinyTim}).

We create a master PSF for each of the F435W, F606W, F814W and F105W filters of the HFF. Since the PSF size is not the same for each filter, we homogenize the process and create PSFs within a radius $r\sim19.5\arcsec$. This is the maximum possible size for F435W with \texttt{TinyTim}, and is the smallest of all four filters. The PSFs are then rotated to match the camera angle, and re-sampled to the pixel size of the HFF images (0.06\arcsec). The images are then deconvolved from the PSF with a Richardson-Lucy algorithm \citep{Richardson1972,Lucy1974}.

\subsubsection{Masking and wavelet processing}
\label{masks}

Bright foreground stars are a major problem, as \texttt{DAWIS} cannot reconstruct them properly, making the masking of these objects mandatory before applying any wavelet process. Otherwise, the strong signal of these objects is found everywhere in the wavelet convolution, contaminating every scale and preventing any type of reliable detection and reconstruction of objects. While the PSF deconvolution is helping in this instance by removing most of the bright components, a few obvious star residuals remain, due to the fact that we did not take into account every variable while computing the PSF with \texttt{TinyTim}, such as spatial variation on the image. 

Here the meaning of the term `mask' is a bit ambiguous, as we do not simply set values covered by the mask either to 0 or to NaN values. Indeed, the wavelet convolution requires a value for each pixel of the image (making the use of NaN values prohibited), and large regions of 0 values would interact strangely with other features of the image, creating artifacts and ghost objects. Since wavelets act as filters in the Fourier space, we replace the values of the masked pixels by random Gaussian noise values, using $\sigma_{\rm{bkg}}$, and the same mean value of standard deviation that we computed in section \ref{sec:background}. The underlying idea is that a mask for a wavelet convolution is a region for which you do not want any correlation with the rest of the image - which is by definition Gaussian noise.

Other sources that are sometimes bringing additional complications are very large foreground elliptical galaxies. The cores of those objects are several orders of magnitude brighter than their halos, making the two components hard to recognize as a single object from the wavelet perspective. For elliptical galaxies at the cluster redshift, the tricky part is to determine at what point the light stops belonging to the galaxy, and starts becoming ICL, but the wavelet convolution is very efficient in decoupling the two components, as they belong to two different luminosity scales. However, for foreground galaxies, this becomes a problem if the luminosity of such a halo merges perfectly with the ICL, both covering the same ranges of surface brightness, as a result of the foreground galaxy being closer to us. In such cases, it is almost impossible to determine the real origin of the light at large radius, and \texttt{DAWIS} cannot differentiate the external parts of this halo from ICL.

In the case of MACS~J0717, a very large and bright foreground galaxy is situated close to the BCG in the projected sky plane (see Fig.~\ref{fig:HFF}, the large yellow elliptical galaxy in the bottom left corner), and the light emitted by this galaxy occupies a very large part of the image. Here we dedicate a specific wavelet treatment to this object. We extract a patch around the galaxy enclosing the full halo and its low surface brightness parts, and run \texttt{DAWIS} in a mode where we only reconstruct detected objects that are centered on the galaxy itself. The idea is to model the galaxy light profile down to very low surface brightness (ICL-like scales), and to remove it from the original FF image, before applying \texttt{DAWIS} to the whole field. We push the number of wavelet detections, reconstructions and subtractions up to $\sim$5-8 consecutive runs (depending on the filter), in order to model precisely and remove as much light as possible, and obtain a map of residuals of mean and standard deviation comparable to the background computed in \ref{sec:background}.

There are two downsides to this method. The first one is that the external parts of the luminosity profile detected and modeled by \texttt{DAWIS} for this galaxy could be amplified by ICL belonging to MACS~J0717. However, we prefer to slightly over-subtract the ICL of MACS~J0717 than to have this strong source of contamination in the results of our study. The second downside is that the extraction of the profile is time-consuming, considering we are running \texttt{DAWIS} several times for a single galaxy, so such a treatment should be restricted to critical cases such as this one.

Once all problematic objects or residuals have been masked or removed, the core wavelet processing of the whole field can start. In order to save some computer time, we re-sample every image to a 0.24\arcsec\ per pixel scale, which is a flux conservative process. We then apply the following wavelet process to the four filters (F435W, F606W, F814W, and F105W) of the HFF.

\begin{figure*}
    \centering
    \includegraphics[width=19cm]{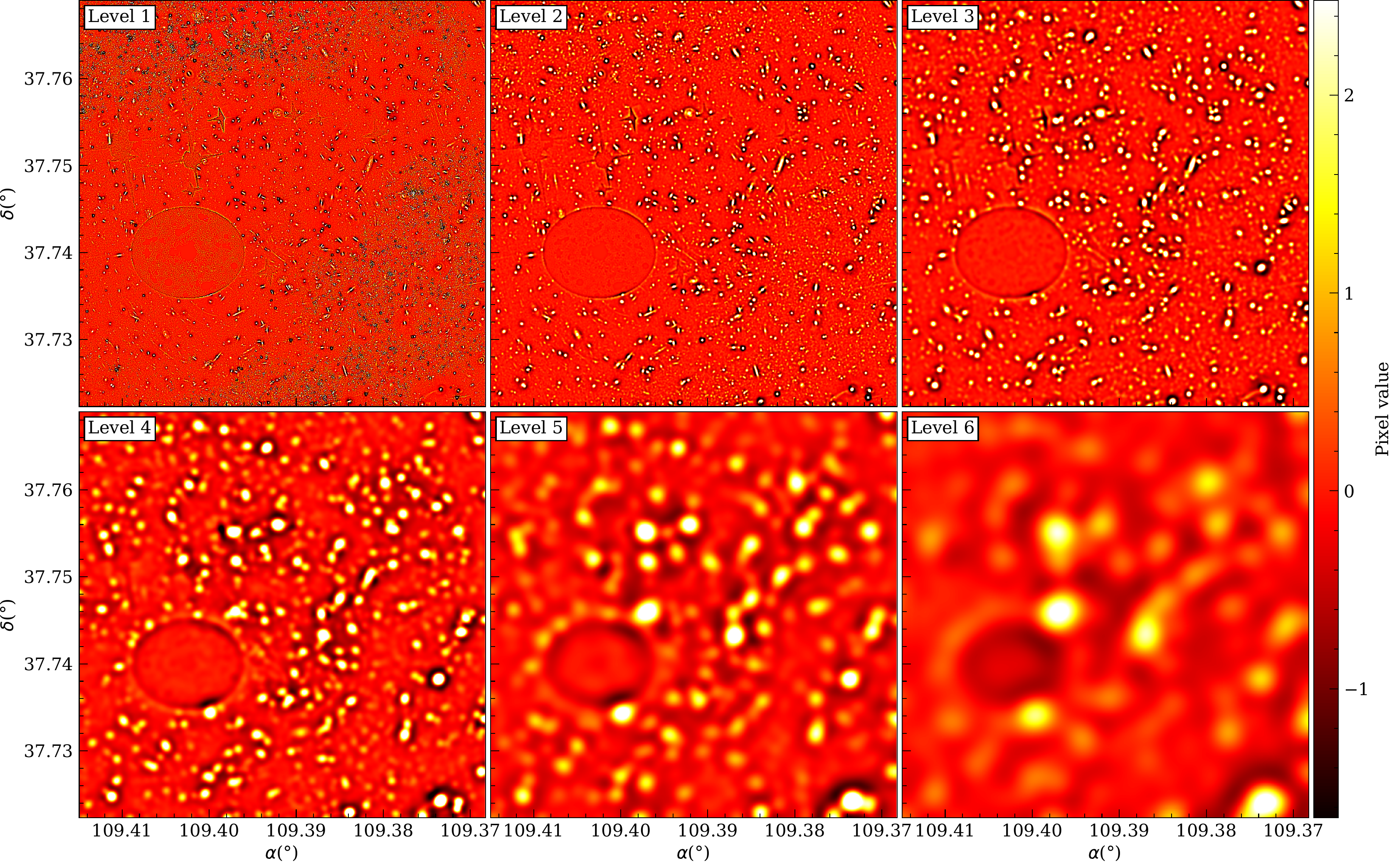}
    \caption{Wavelet planes of the first run of \texttt{DAWIS} on the F814W image. There are six planes, corresponding to maximum characteristic object sizes of 64 pixels. As the level of the plane increases, the size of sources becomes larger and their intensity decreases. When the thresholding is done, the negative coefficients are simply ignored.}
    \label{WP}
\end{figure*}

\begin{enumerate}

    \item A first run of \texttt{DAWIS} with a very high wavelet detection threshold ($k=10$) and a small convolution kernel ($N_{\rm{lvl}}$=6, corresponding to a characteristic object size of 64 pixels), in order to detect bulges and other very bright and localized sources. The wavelet planes of the F814W image can be found in Fig.~\ref{WP} as an example illustrating this step;
    
    \item A second run of \texttt{DAWIS} with a lower wavelet detection threshold ($k=5$) and a larger kernel ($N_{\rm{lvl}}$=7, corresponding to a characteristic object size of 128 pixels), in order to detect disks and outer halos of galaxies;
    
    \item A third run with an even lower detection threshold ($k=3$) and with $N_{\rm{lvl}}$=9 (the maximum value given the size of the images), which gives a characteristic object size of 512 pixels. Extended halos that have been missed in the second run can be detected this way.
    
\end{enumerate}

\begin{figure*}
    \centering
    \includegraphics[width=15cm]{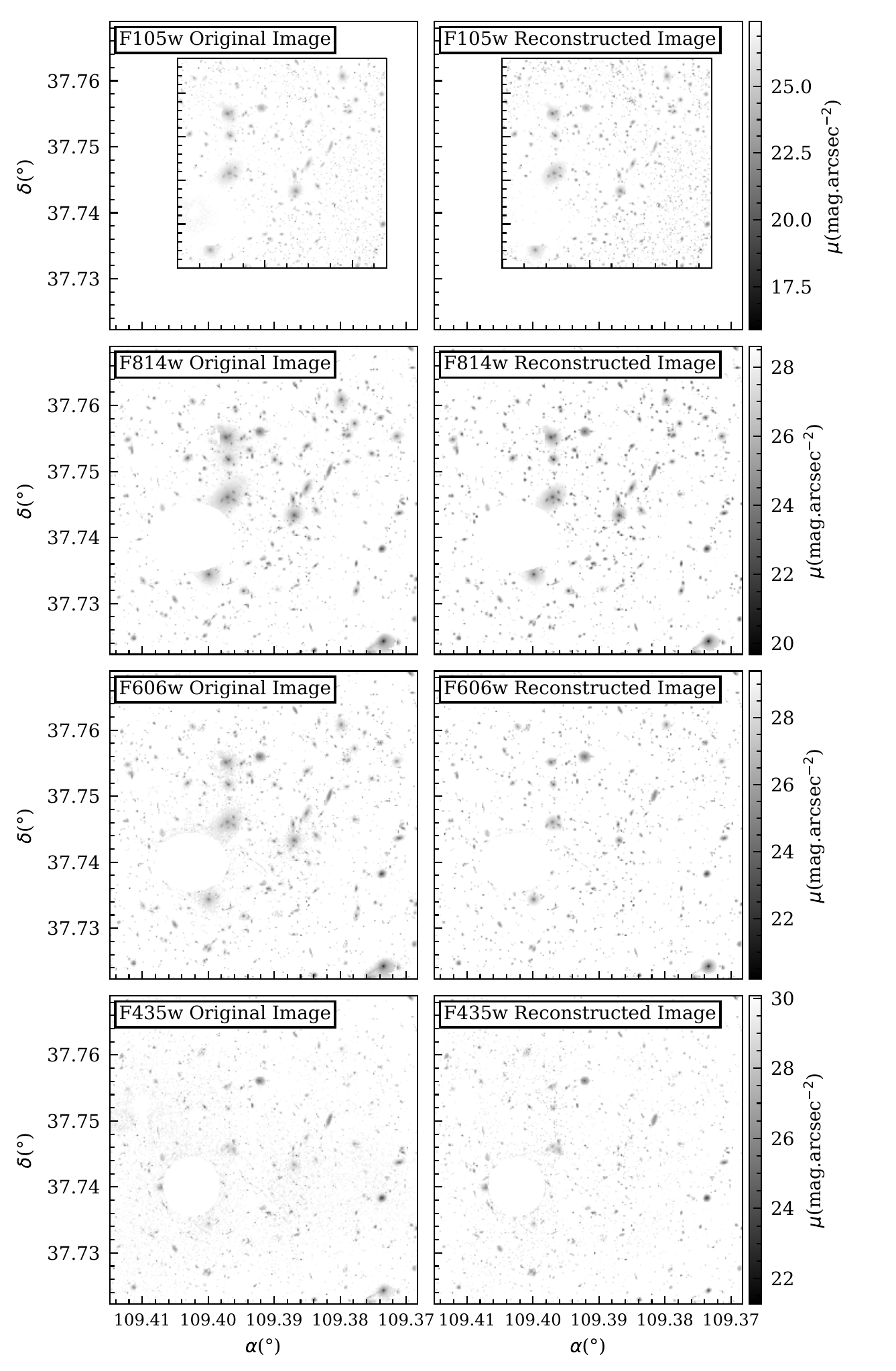}
    \caption{Left column: original HFF images in the four filters after PSF deconvolution and masking of star residuals and of the large foreground galaxy. Right column: Stacked images of objects detected and reconstructed by three runs of \texttt{DAWIS}. The F105W residual map is  smaller as a result of the WFC3 field of view being smaller than the ACS one, and has been scaled to the other ones in consequence.}
    \label{Recim}
\end{figure*}

A complete reconstructed image of every object detected by \texttt{DAWIS} can be created by stacking the reconstructed images of the three runs (see Fig.~\ref{Recim}). There is an excellent agreement between the original and reconstructed images, which demonstrates the ability of DAWIS to separate compact objects (such as galaxies) from more extended sources (the ICL) that remain in the residual image.

\subsubsection{Results}
\label{sec:resultsICL}

\begin{figure*}
    \centering
    \includegraphics[width=19cm]{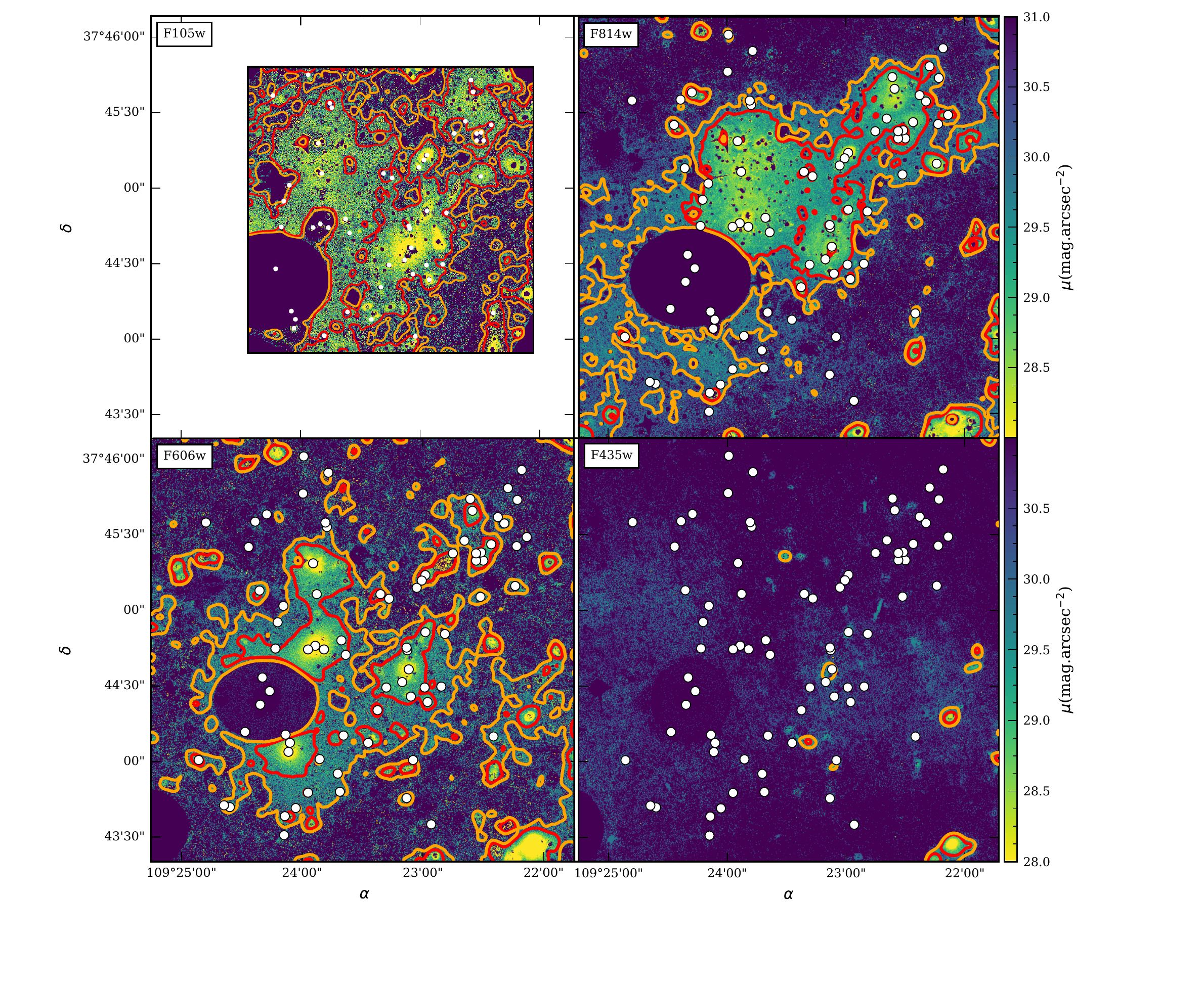}
    \caption{Surface brightness maps of the residuals after wavelet processing by \texttt{DAWIS} in each band. The orange contours show 3$\sigma _{\rm bkg}$ detection and the red ones 5$\sigma _{\rm bkg}$.
    The white dots show the galaxies in the cluster redshift range ($0.53<z<0.56$). From top to bottom and left to right : F435W, F606W, F814W, F105W. The F105W residual map is  smaller as a result of the WFC3 field of view being smaller than the ACS one, and has been scaled to the other ones in consequence. The contours are smoothed with a Gaussian kernel of $\sigma$=5 for the map to be readable.}
    \label{mapICL}
\end{figure*}

To create the ICL maps in the HFF, we choose to use the residual images after the three \texttt{DAWIS} runs. For each filter, we attribute pixels in the residual maps with a value above 3$\sigma_{\rm{bkg}}$ to the ICL ($\sigma_{\rm{bkg}}$ is the global standard deviation of the sky background computed in \ref{sec:background}). The final ICL maps are given in Fig.~\ref{mapICL}, showing 3$\sigma_{\rm{bkg}}$ and 5$\sigma_{\rm{bkg}}$ contours.

The spatial distribution on the projected sky of the ICL in each filter is consistent with previous works on MACS~J0717 such as in \citet{Morishita2017} (see their Figure 2) or in \citet{Montes2019} (see their Figure 3). The fact that the morphology of the ICL differs from one filter to another is interesting, as it indicates the presence of different populations of stars in the ICL.

We retrieve galaxy spectroscopic redshifts from the NASA Extragalactic Database\footnote{https://ned.ipac.caltech.edu/} (NED) and select only galaxies in the FoV of the HFF and with $ 0.53 < z_{\rm{spec}} < 0.56$, assuming galaxies in this range of redshift are associated with MACS~J0717. We add the position of the spectroscopically confirmed cluster member galaxies on the residual maps. We note that while this galaxy catalogue is good for representation, it is far from being complete. Instead, we use the red sequence (RS) computed in Section~\ref{sec:GLF} to compute ICL fractions in each band:

$$
f_{\rm{ICL}}=\frac{F_{\rm{ICL}}}{F_{\rm{gal}}+F_{\rm{ICL}}}
$$

\noindent
where $F_{\rm{ICL}}$ is the integrated flux of ICL, and $F_{\rm{gal}}$ the integrated flux of the galaxies belonging to MACS~J0717. $F_{\rm{ICL}}$ is obtained by summing the pixel values with values greater that 3$\sigma_{\rm{bkg}}$ in the ICL maps, and $F_{\rm{gal}}$ by summing the pixel values of the RS galaxy profiles in the reconstructed images. In order to compare our approach with recent works, we measure the ICL fractions in the same radii centered on the BCG than in \citet{Jimenez-Teja2019}. The computed ICL fractions can be found in Table~\ref{tab:detectionlimit}. 

The errors on the ICL fraction are computed with a bootstrap on the values of the pixels of the galaxies and of the pixels of the ICL. For each filter, we create a sample with all the pixels belonging to RS galaxies of the reconstructed image, and a sample with all the ICL pixels of the residual image. We draw $N=10000$ sub-samples randomly from each sample, allowing the same pixel value to be drawn multiple times. We then compute the given ICL fraction for each sub-sample which gives $N$ values of ICL fractions for each filter. The errors on the true ICL fraction value are then estimated by computing a 95\% confidence interval on the sub-sample values.

We find ICL fractions in good agreement with the ones in \citet{Jimenez-Teja2019} in the F606W and F814W filters. Our fraction in the F435W filter differs significantly (we find a value of 2.5\% compared to their value of 7.22\%), which can be due to a number of differences in the data processing, since the sky level is computed differently in both studies, and to the fact that these authors used a different analysis package \texttt{CICLE} \citep[CHEFs Intracluster Light Estimator;][]{Jimenez-Teja2016} to extract and remove galaxy light profiles. In the case of F105W (not analysed by \citet{Jimenez-Teja2019}), we simply integrated the whole FoV to compute the fraction. We find that the fraction of ICL is peaking in the blue F606W filter before decreasing progressively in the F814W and F105W filters as it gets redder.

\subsection{Detection of IFL in the filament}
\label{sec:tidalstreams}

In this section, we look for IFL in the regions corresponding to the filament of MACS~J0717 (regions B and C, see Fig.~\ref{fig:HSTmosaic} for a global view of the system). We first investigate if we detect a global diffuse component in the filament, as in the core of MACS~J0717, then we look at perturbed systems of galaxies in the filament presenting evident tidal streams.

\subsubsection{Detection of diffuse sources}

Since the HST mosaic is not as deep as the HFF, we test the feasibility of the detection of ICL in those images. For this, we try to re-detect in the corresponding HST images the ICL previously found in the core of MACS~J0717 (see Section~\ref{ICLcore}). After binning the image from 0.05\arcsec\ to 0.24\arcsec\ per pixel, the stars and the foreground galaxy are masked by hand. We do not deconvolve the image by the PSF, since the result is not used in any way other than testing the detection limit of the mosaic. Contrary to the HFF, there is no parallel field to compute the sky background level, but since we now know the spatial distribution of the ICL, we use the same method as in Section~\ref{sec:background}. This time though, the regions are created directly in the image used for the detection, avoiding the areas where ICL is detected in the HFF. We then run exactly the same wavelet process as in Section~\ref{masks} on the whole image. This gives a 3$\sigma_{\rm bkg}$ detection threshold of 28.54 mag.arcsec$^{-2}$ for the F814W image, and 27.59 mag.arcsec$^{-2}$ for the F606W image. 

\begin{figure}
  \centering
  \includegraphics[width=3.5in,clip=true]{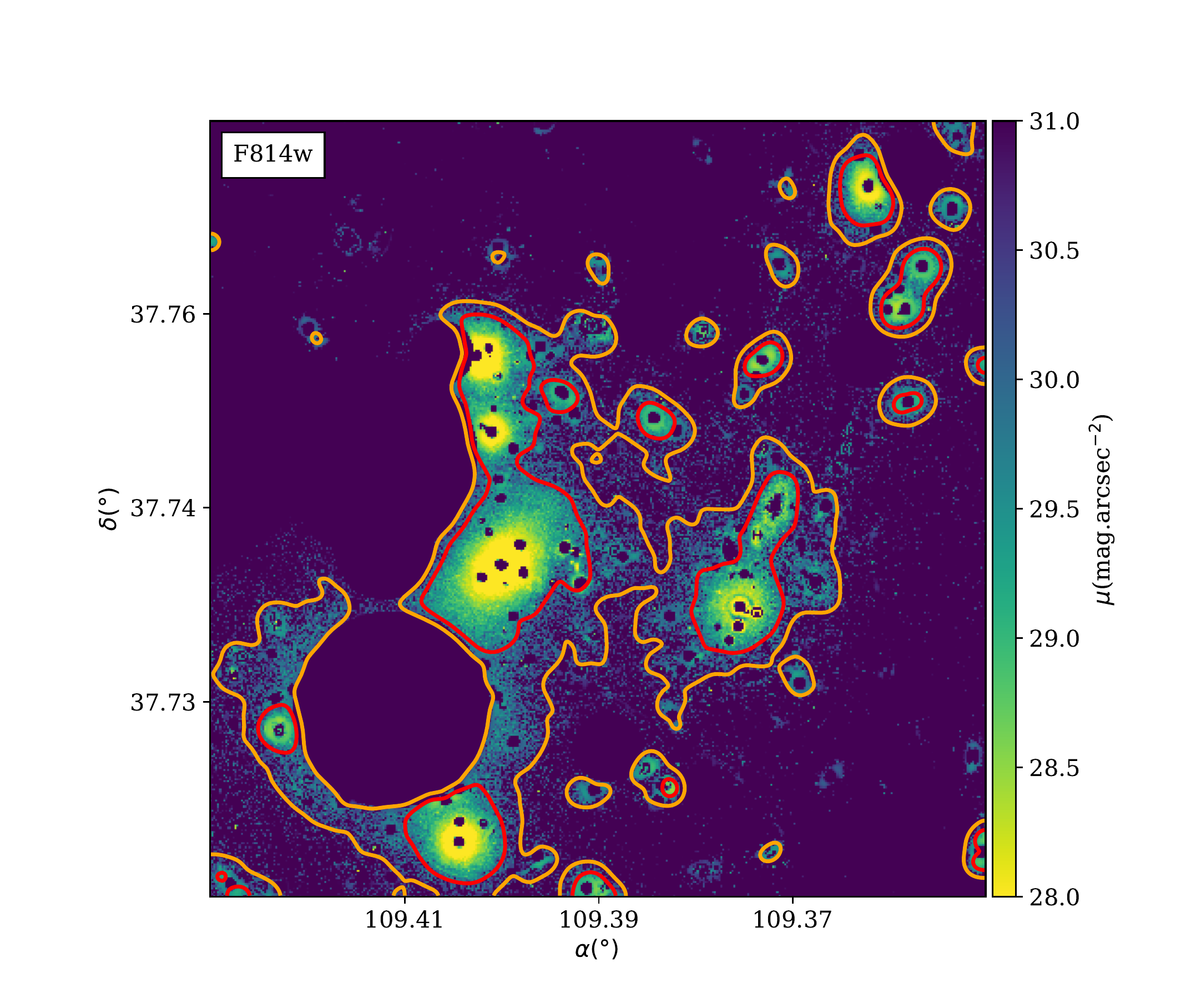}
  \caption{Close-up on the ICL in the core of MACS~J0717 in the HST mosaic F814W image. 
  The stars and the large foreground galaxy are masked by hand (the external halo of the galaxy is still visible here), and the same wavelet processing than for the HFF has been applied. The contours show the 3$\sigma_{\rm{bkg}}$ detection limit. Part of the ICL contribution is detected in this image, showcasing the fact that this mosaic is deep enough to detect the brightest component of the ICL.}
  \label{fig:f814_mathilde}
\end{figure}

The test is not conclusive in the case of the F606W image which is too shallow to reach the surface brightness level necessary to detect ICL, and as the residual image shows many factors of contamination due to flat-fielding uncertainties. In the F814W residual image though, we are able to detect once again ICL in the core of MACS~J0717 (see Fig.~\ref{fig:f814_mathilde}), meaning that the mosaic in the F814W filter is deep enough to detect large sources of equivalent surface brightness that could be associated to IFL in the filament. We apply the same process to every image of the HST mosaic covering parts of the filament in the F814W filter. If possible, the background regions are created in areas outside of the contours demarcating regions B and C shown in Fig.~\ref{fig:HSTmosaic}. If not, they are created at random positions in the image.

 We do not detect any significant source of light that could be associated to IFL, as the residual maps mostly contain noise, flat-field uncertainties and wavelet residuals. This is not really surprising, as strong diffuse light features are associated to galaxy clusters or groups (respectively ICL and IGL) and are believed to be formed through many galaxy-galaxy gravitational interactions such as mergers. This absence of IFL detection seems to confirm this formation scenario, as cosmic filaments do not undergo as many gravitational interactions as galaxy groups or clusters (in our case, interactions are not sufficient to create an amount of IFL equivalent to the amount of ICL in the core of MACS~J0717).

\subsubsection{Detection of tidal streams}

In parallel, we look for tidal streams in the cosmic filament of MACS~J0717, as simulations have shown that significant portions of ICL could be formed in such structures \citep{Rudick2009}. The detection of tidal streams is a tricky task due to their various morphologies, and while some could be found in the \texttt{DAWIS} residual images, we can't really differentiate them from wavelet residuals and artifacts. We choose instead to detect the presence of candidate tidal streams by visually inspecting every image of the \emph{HST} mosaic. The work in this section is only qualitative and meant to be a complement to the study of the ICL in the core of MACS~J0717, and a pinpoint for future studies.

We look first at every galaxy with a spectroscopic redshift in the range [0.53,0.56], then at every galaxy of the RS computed in Section~\ref{sec:GLF}. The criteria used to determine the tidal stream candidates are the following: i) presence of several galaxies with close positions on the projected sky plane; ii) signs of very disturbed galaxy morphology; iii) presence of low surface brightness features such as tidal streams or arcs associated with these galaxies. Several inspections were independently done by different persons of our group before comparing the results and picking the candidates.

\begin{figure}
  \centering
  \includegraphics[width=3.5in,clip=true]{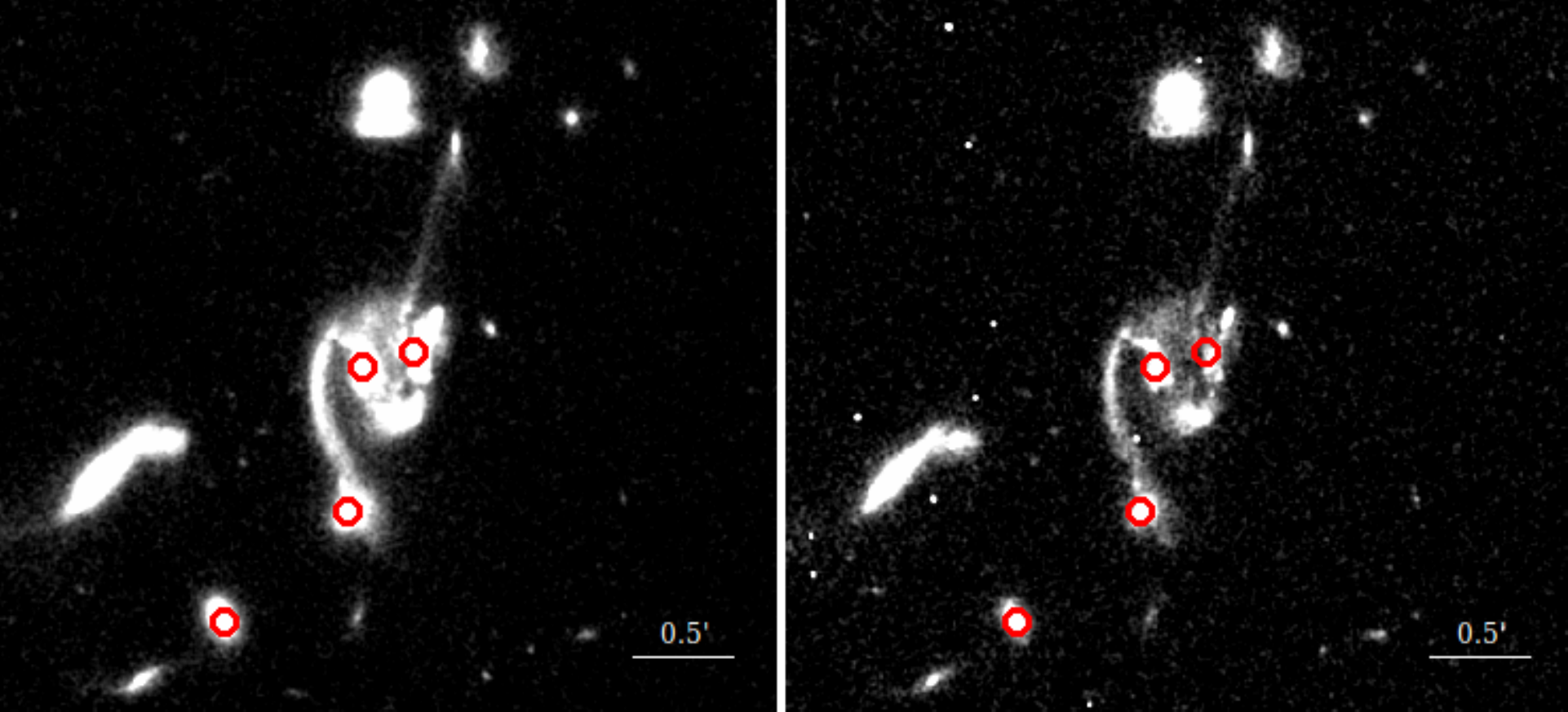}
  \caption{Close-up on the disturbed system of galaxies in the region B of the cosmic filament. Left: F814W filter, right: F606W filter. The red circles mark galaxies with spectroscopic redshifts in the range [0.53,0.56].}
  \label{fig:cygne}
\end{figure}

In the whole mosaic we find only one system matching these criteria, in region B of the filament, and selected from the spectroscopic redshift catalogue. The system is composed of three interacting galaxies with spectroscopic redshifts, presenting obvious signs of tidal stripping, such as a linear stream, an arc, and a diffuse envelope around the core of the galaxies (see Fig,~\ref{fig:cygne}). This system resides within the weak lensing contours of \citet{Martinet2016} (see Fig.~\ref{fig:HSTmosaic}), indicating that it belongs to a massive substructure, such as a galaxy group embedded in the cosmic filament (see also Sections~\ref{sec:GLF} and \ref{sec:PAs}), and producing its own IGL through tidal streams. The fact that the only system of galaxies presenting obvious tidal streams in the whole cosmic filament seems to actually belong to this galaxy group suggests that the creation of IFL through galaxy-galaxy mergers directly in the filament (e.g. outside of a massive sub-structure) is not possible, at least not in a significant way.

\section{Galaxy luminosity functions}
\label{sec:GLF}

This section aims at better understanding the distribution of galaxy luminosities within filaments. This is done by measuring the galaxy luminosity functions (GLFs) in the structures surrounding MACS~J0717 and comparing them with the cluster and field GLFs.

\subsection{Computing GLFs}
\label{subsec:compuGLF}

The building of the GLFs follows that of \citet{Martinet2017} and we refer the reader to this paper for a detailed description of each step. Here we only recall the salient points of the analysis and those which differ from the mentioned study.

The detection of objects is made individually on each image of the \emph{HST} mosaic with the \texttt{SExtractor} software \citep{Bertin1996}, and the catalogs are then concatenated in a single catalog for each band. Because the cluster core and the mosaic covering the filament were observed at different epochs and with different observing strategies, the astrometry between the two sets of images does not match with a sufficient accuracy. This results in some objects in the overlapping area between these images being detected several times. For each of these objects we discard the lowest signal-to-noise detection applying a matching of objects in separation, magnitude, and surface brightness, in the mentioned area. Due to the same astrometric issue, the detection is made independently in the F814W and F606W bands, contrary to \citet{Martinet2017} where we used the double image mode of \texttt{SExtractor} to measure the flux in the F606W filter in the same apertures as those detected in the F814W filter. In the present study, the two catalogs are then cross-matched in a closest neighbor approach, with a maximum separation of 2\,arcsec. We also discard spurious or contaminated detections using hand-made masks around bright saturated stars and on noisy image edges. All the magnitudes discussed hereafter correspond to \texttt{SExtractor} {\small MAG\_AUTO} measurements.

Galaxies are then separated from stars based on a maximum surface brightness versus magnitude diagram, in the F814W filter up to magnitude 25.

We select RS galaxies in a color magnitude diagram through an iterative process. We first consider a broad RS centered on the F814W$-$F606W color of elliptical galaxies at the cluster redshift applying prescriptions from \citet{Fukugita1995}. The RS is then refined with a linear fit to the selected galaxies by fixing the slope to $-0.0436$, which has been shown to be a constant value for clusters in this redshift range \citep{Durret2011}. The final width of the RS is set to $\pm0.3$ in color.

We subtract field galaxies in apparent magnitude after applying the same RS color cut as that of the cluster galaxies and normalizing both cluster and field galaxy counts to 1~deg$^2$. The sample of field galaxies we used is that of \citet{Martinet2017}, which corresponds to a subarea of $\sim0.05$~deg$^2$ of the COSMOS survey\footnote{\url{http://cosmos.astro.caltech.edu/}} re-analysed by the 3D-HST team \citep{Brammer2012,Skelton2014}, and for which we applied the same detection setup as in the present study.

We compute restframe absolute magnitudes for RS galaxies by applying the distance modulus at the cluster redshift and a constant k-correction for all galaxies. This assumes that RS galaxies lie at the same redshift and have identical colors, which is a generally good approximation for RS galaxies \citep[see e.g.][]{Martinet2017}. The k-correction is computed as the mean over a series of early-type spectral energy distribution templates in a $\pm0.05$ redshift range around the cluster redshift using the \texttt{LePhare} software \citep{Arnouts1999,Ilbert2006}.

The last steps assume that galaxies in the filaments also lie in the cluster RS, an assumption that will be discussed when interpreting the results. We note that in the present study it is not possible to consider photometric redshifts, since we only have two optical bands in the region covered by the filament.

Each magnitude bin is assigned an error bar which corresponds to the quadratic sum of the Poisson errors on the field, and on the cluster or filament counts. 
Given the very deep images of MACS~J0717, the completeness limit is set by the depth of the COSMOS field galaxies, determined from their magnitude histogram. These limits correspond to a magnitude of 26 in both F814W and F606W filters.

Finally, we fit Schechter functions \citep{Schechter1976} to the GLFs:
\begin{equation}
\label{eq:schech}
 N(M)  = 0.4\ln(10)\phi^*[10^{0.4(M^*-M)}]^{(\alpha+1)}\exp(-10^{0.4(M^*-M)}),
\end{equation}
constraining the three following parameters: the slope of the faint end $\alpha$, the characteristic magnitude of the bending $M^*$, and the normalization $\phi^*$. The fit is performed via a $\chi^2$ minimization.


\subsection{Red sequence GLFs}

\begin{figure*}
  \centering
  \begin{tabular}{cc}
  \includegraphics[width=3.5in,clip=true]{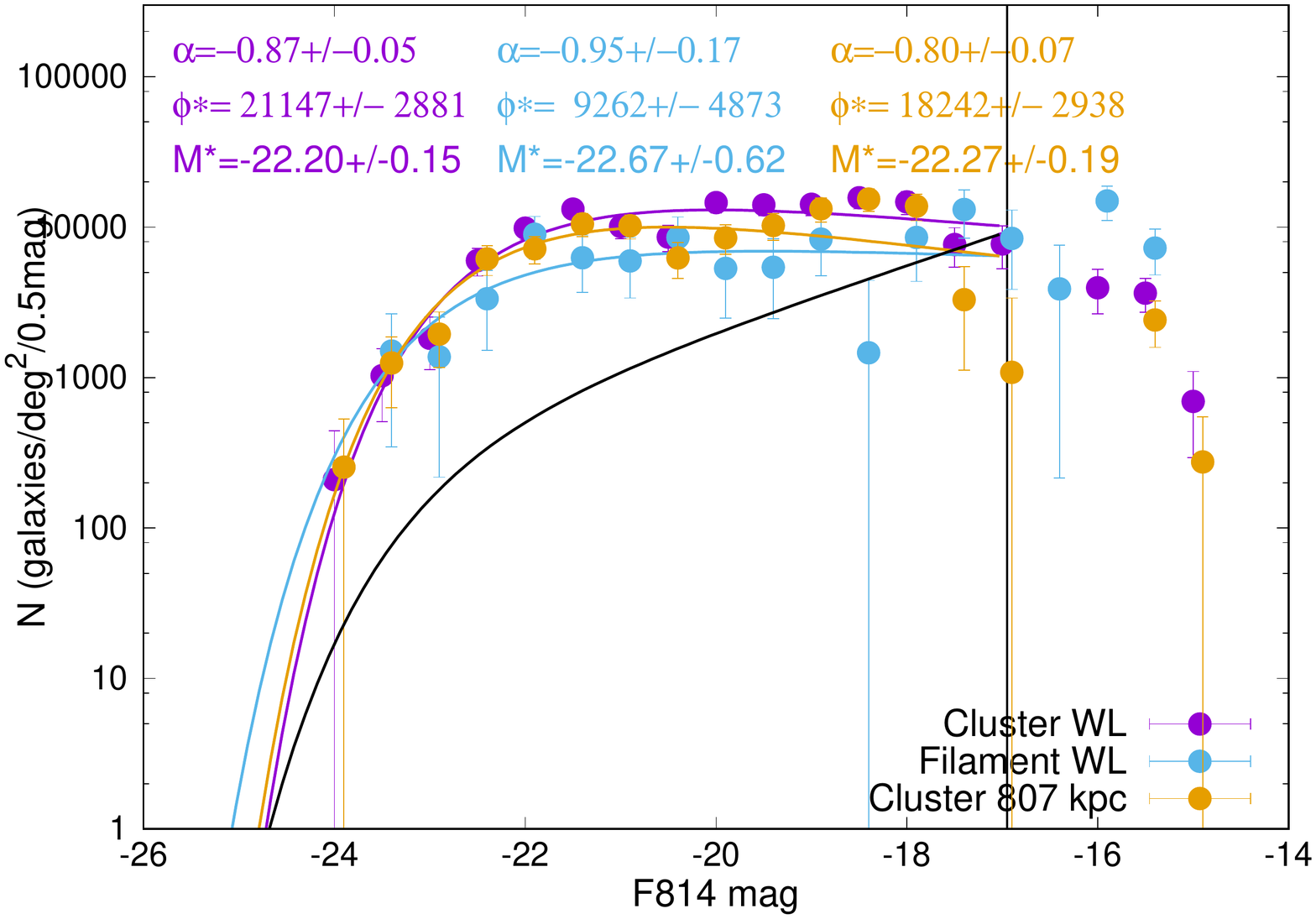}&\includegraphics[width=3.5in,clip=true]{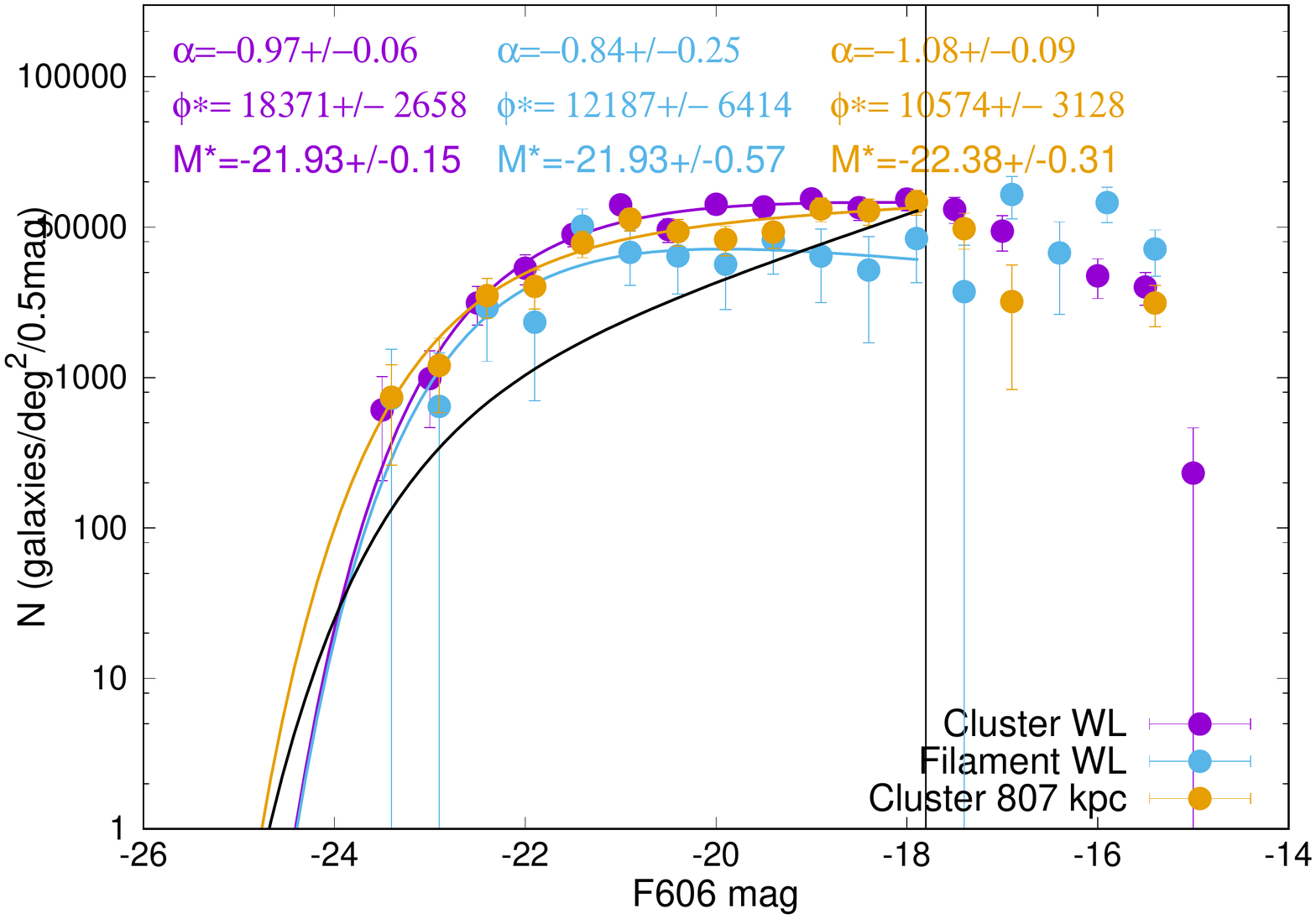}\\
  \includegraphics[width=3.5in,clip=true]{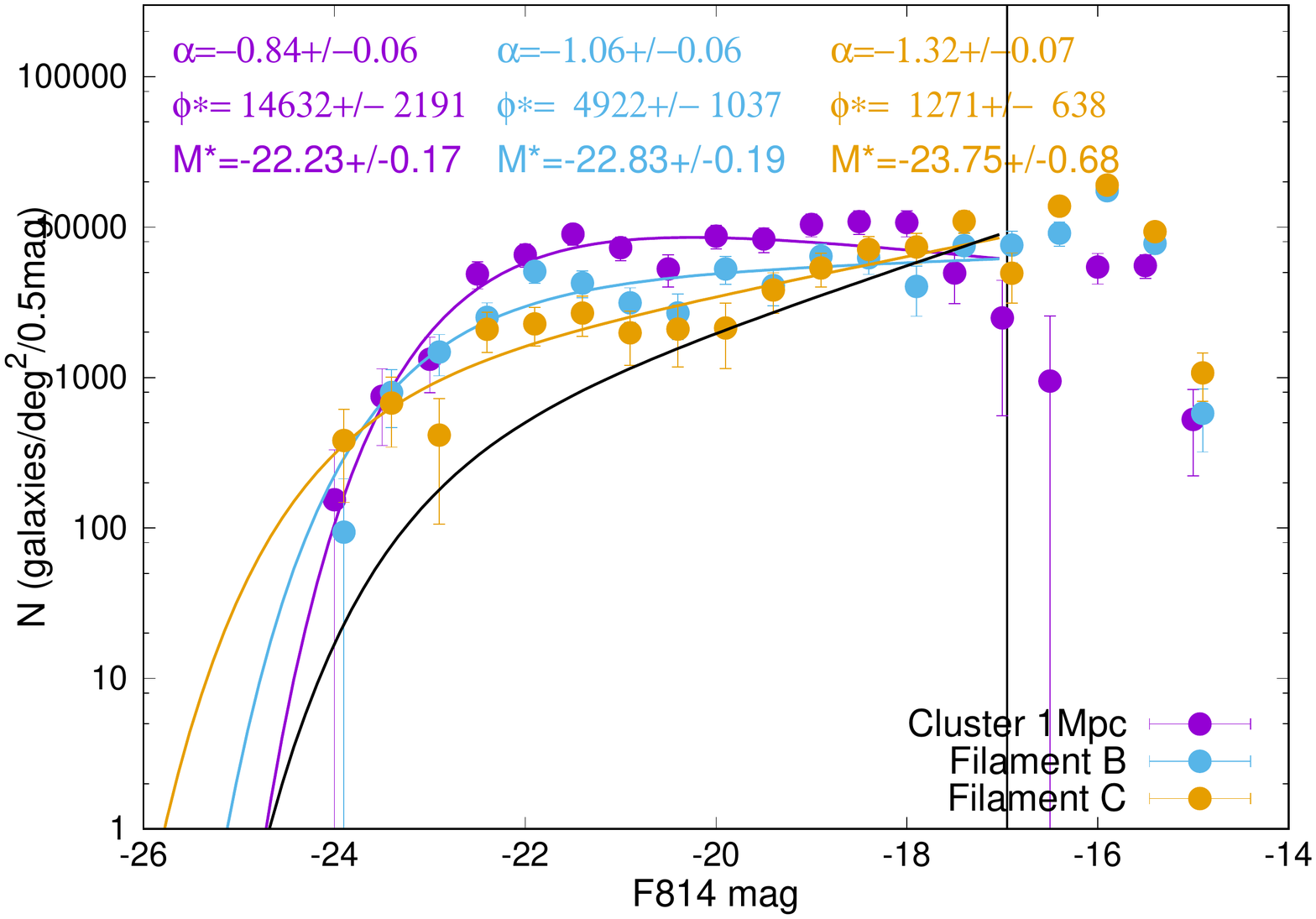}&\includegraphics[width=3.5in,clip=true]{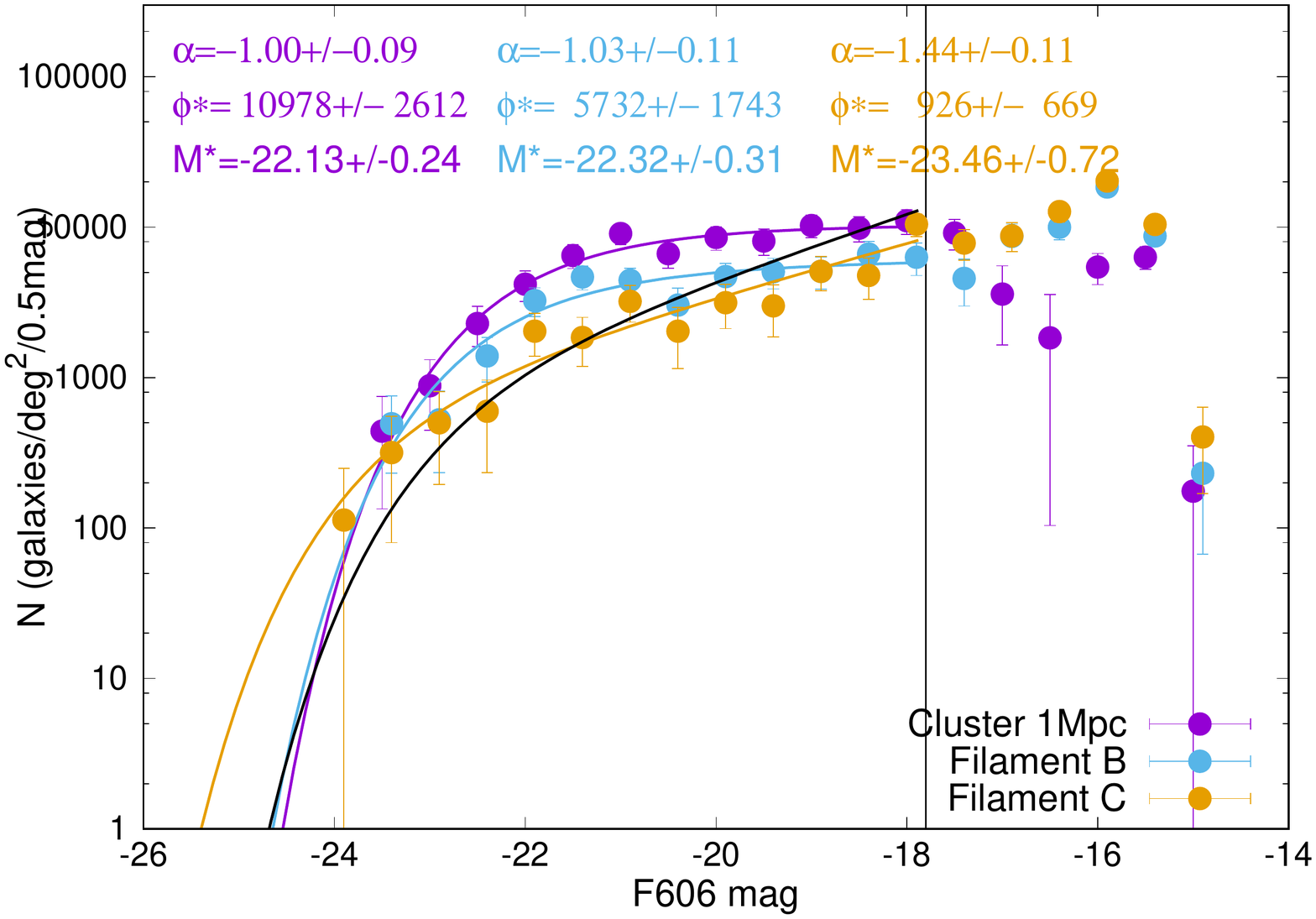}\\
  \end{tabular}
  \caption{Red sequence GLFs in the F814W (\textit{left}) and F606W (\textit{right}) filters}. \textit{Top} shows the GLFs centered on the cluster in a 807~kpc radius (yellow), of the cluster WL contours (violet), and of the southeastern filament WL contours (cyan). \textit{Bottom} shows GLFs centered on the cluster in a 1~Mpc radius (violet), and in the optically detected filaments B (cyan) and C (yellow). The different curves are the Schechter fit to the data, and the parameters of the fit are displayed in the same color. In both plots the black curve corresponds to the GLF that we would observe if we were selecting field galaxies and wrongly assume that they lie at the cluster redshift. The black vertical line corresponds to the completeness magnitude limit.
  \label{fig:glf1}
\end{figure*}

We compute the GLFs in six different subareas of the \emph{HST} mosaic. First, we study the GLF in a region within a 807~kpc radius from the cluster center. This is the same area as the one studied in \citet{Martinet2017} for the same cluster with \emph{HST} data, and we study it here to verify that our analysis is consistent with previous ones. Second, we measure the GLFs in the $3\sigma$ weak lensing (WL) contours shown in Fig.~\ref{fig:HSTmosaic}
(as in \citet{Martinet2016}). These contours correspond to the total overdensity, both luminous and dark matter, as probed by the lensing of background sources by the gravitational potential of the foreground structures. \citet{Martinet2016} reported an almost $11\sigma$ detection of the cluster itself and a $\sim8\sigma$ detection of a filamentary structure to the southeast from Subaru/Suprime-Cam images, which was also detected in the WL analysis of the \emph{HST} mosaic used in the present paper \citep{Jauzac2012}. Finally, we compute the GLFs in a 1~Mpc radius centered on the cluster, and in two ellipses defined from the overdensities of luminous RS galaxies after filtering the galaxy density field with a Gaussian kernel. The latter areas are defined in Fig.~\ref{fig:HSTmosaic} (labelled as `filament B' and `filament C' in Figure\,5 of \citet{Durret2016}. We note that the filament detected from WL is included in `filament B', and that `filament C', lying to the south of the previous structure, is only weakly detected with lensing compared to the other structures \citep[the $3\sigma$ detections in][]{Martinet2016}.

The RS GLFs for the six different areas are presented in Fig.~\ref{fig:glf1}, in the F814W and F606W bands, and the parameters of the Schechter fits are summarized in Table~\ref{tab:schech}. The different subareas are populated with 1127, 735, and 1035 RS galaxies for ``Cluster 1Mpc'', ``Cluster 807kpc'', and ``Cluster WL'' respectively, and with 254, 1640, and 1423 galaxies within the RS color cut for ``Filament WL'', ``Filament B'', and ``Filament C'' respectively. The GLFs are however normalized to 1 deg$^2$. Since we have only two optical bands, we assume that the RS cut is a good selection of galaxies lying at the cluster redshift. We will relax this assumption later in this section.



\begin{table*}
\centering
  \caption{Schechter fit parameters for RS GLFs for the six different areas. $\alpha$, $M^*$, and $\phi^*$ correspond to the faint end slope, the characteristic magnitude, and the normalization, respectively.}
  \begin{tabular}{l|ccc|ccc}
    \hline
    \hline
    &&F814W&&&F606W&\\
    \hline
    &$\alpha$&$M^*$&$\phi^*$&$\alpha$&$M^*$&$\phi^*$\\
    &(dimensionless)&(mag)&(galaxies.deg$^{-2}$)&(dimensionless)&(mag)&(galaxies.deg$^{-2}$)\\
    \hline
    Cluster 807~kpc&$-0.80\pm0.07$&$-22.27\pm0.19$&$18242\pm2938$&$-1.08\pm0.09$&$-22.38\pm0.31$&$10574\pm3128$\\
    Cluster 1000~kpc&$-0.84\pm0.06$&$-22.23\pm0.17$&$14632\pm2191$&$-1.00\pm0.09$&$-22.13\pm0.24$&$10978\pm2612$\\
    Cluster WL&$-0.87\pm0.05$&$-22.20\pm0.15$&$21147\pm2881$&$-0.97\pm0.06$&$-21.93\pm0.15$&$18371\pm2658$\\
    Filament WL&$-0.95\pm0.17$&$-22.67\pm0.67$&\,\,\,$9262\pm4873$&$-0.84\pm0.25$&$-21.93\pm0.57$&$12187\pm6414$\\
    Filament B&$-1.06\pm0.06$&$-22.83\pm0.19$&\,\,\,$4922\pm1037$&$-1.03\pm0.11$&$-22.32\pm0.31$&\,\,\,\,$5732\pm1742$\\
    Filament C&$-1.32\pm0.07$&$-23.75\pm0.68$&$1271\pm638$&$-1.44\pm0.11$&$-23.46\pm0.72$&\,\,\,\,$926\pm669$\\
    \hline
  \end{tabular}
  \label{tab:schech}
\end{table*}

We first compare the results of the RS GLF computed in the 807~kpc radius with those of \citet{Martinet2017} for the same cluster. We find a perfect agreement between the two studies in both bands. Quantitatively, we find  faint end slopes $\alpha=-0.80\pm0.07$ and $\alpha=-1.08\pm0.09$ in F814W and F606W, respectively, while \citet{Martinet2017} found $\alpha=-0.84\pm0.44$ and $\alpha=-1.11\pm0.51$. Although the values are almost identical, we note the tremendous gain in precision, explained by the extension of the fit to galaxies almost 4 magnitudes fainter.

The GLF in the cluster WL contours is almost identical to that of the inner part of the cluster, showing that more complex extended contours defined by WL are probing the same galaxy population as that of the cluster core. In the WL contours of the filamentary structure, we note a drop in the galaxy density but the shape of the GLF remains the same. This last result is a hint that the structure selected through its lensing effect is probably a pre-processed group, embedded in the filament, which already contains a RS, and is falling onto the cluster. Although we do not know what a filament GLF looks like, we expect it to lie somewhere between the field and cluster GLFs. We also show the GLFs of field galaxies with the same color selection as the RS, and assuming they lie at the cluster redshift. This last assumption is false but allows us to see what the RS GLF would look like if we misinterpret a field region for a filament. We see that it would give rise to a lower bright galaxy density and to a steeper faint end than what is observed in this potential group of galaxies. The results in the F606W filter for the WL-defined structures are very similar to those in the F814W filter, with a slightly lower number of bright galaxies above the field.

We now look at look at the GLFs defined within the optical contours of \citet{Durret2016}. The cluster in a 1~Mpc radius presents a typical GLF for that redshift, in good agreement with those computed both in a 807~kpc radius, and within the WL contours. The GLF in filament B shows a lower density than that within the WL contours, but also a slight steepening of the faint end, intermediate between the cluster and field behaviors in the F814W filter. This comforts us in our interpretation of the extended structure detected in the optical being a filament with a WL-detected overdensity corresponding to a group within the filament. Filament C has an even steeper faint end, very close to what we would get if it was made of field galaxies, but with a significant overdensity at the bright end in F814W. The results are similar in F606W, although the bright overdensity is less pronounced for filament B, and filament C shows no difference from the case where we select red field galaxies. It is therefore possible that filament C does not correspond to a filament but more probably just to a few red bright galaxies superimposed on the field, a hypothesis which is also suggested by its low WL detection. The differences between the F606W and F814W bands can be understood considering the restframe wavelength that is probed. At a redshift of $z=0.5458$, the F606W filter corresponds to the rest frame blue band, and the absence of bright galaxies in this filter highlights a low star-forming efficiency in the filamentary structures. In F814W however, we observe rest frame wavelengths corresponding to the g and r filters, and see passive galaxies that populate the cluster RS.

\begin{figure*}
  \centering
  \begin{tabular}{cc}
  \includegraphics[width=3.5in,clip=true]{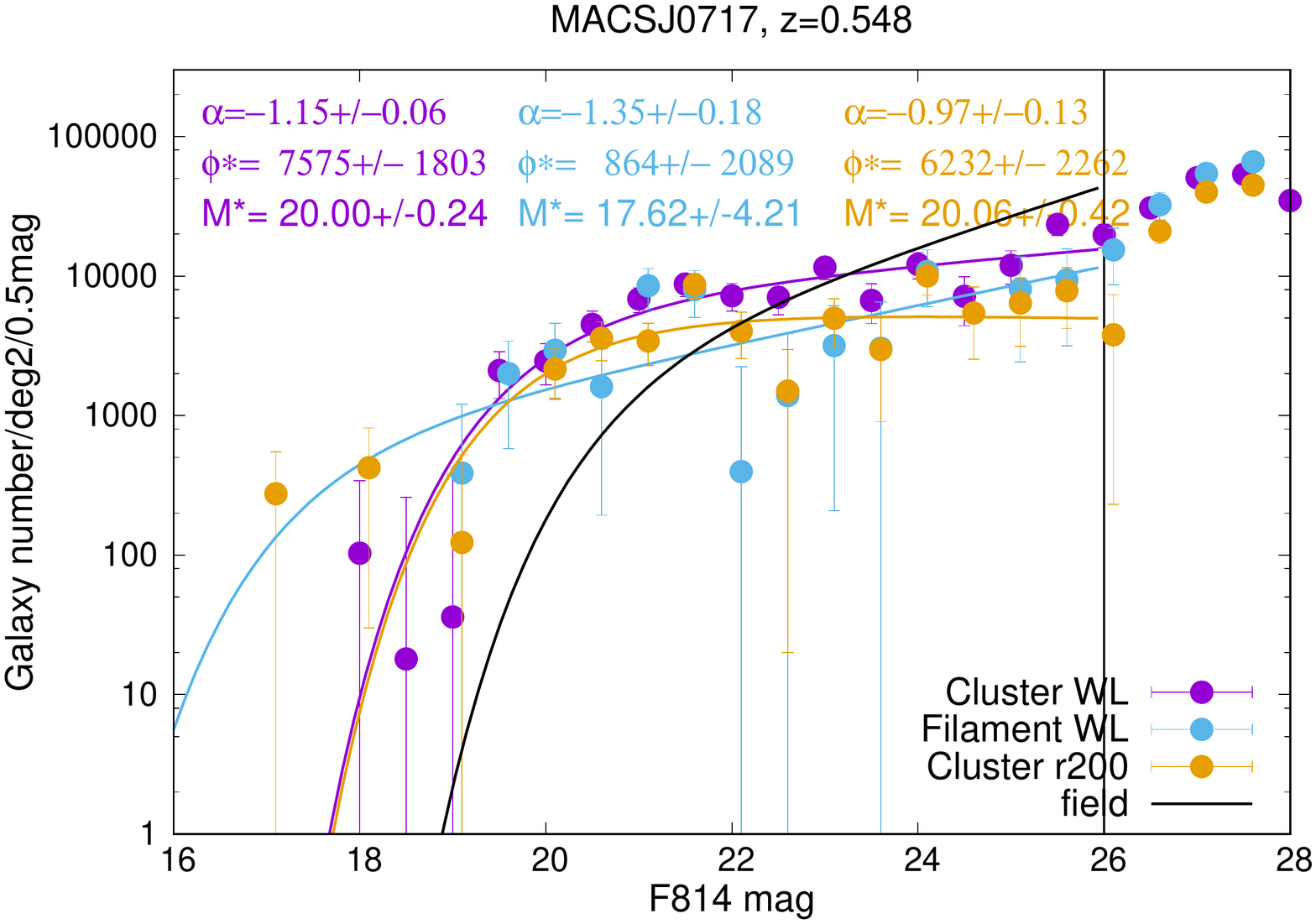}&\includegraphics[width=3.5in,clip=true]{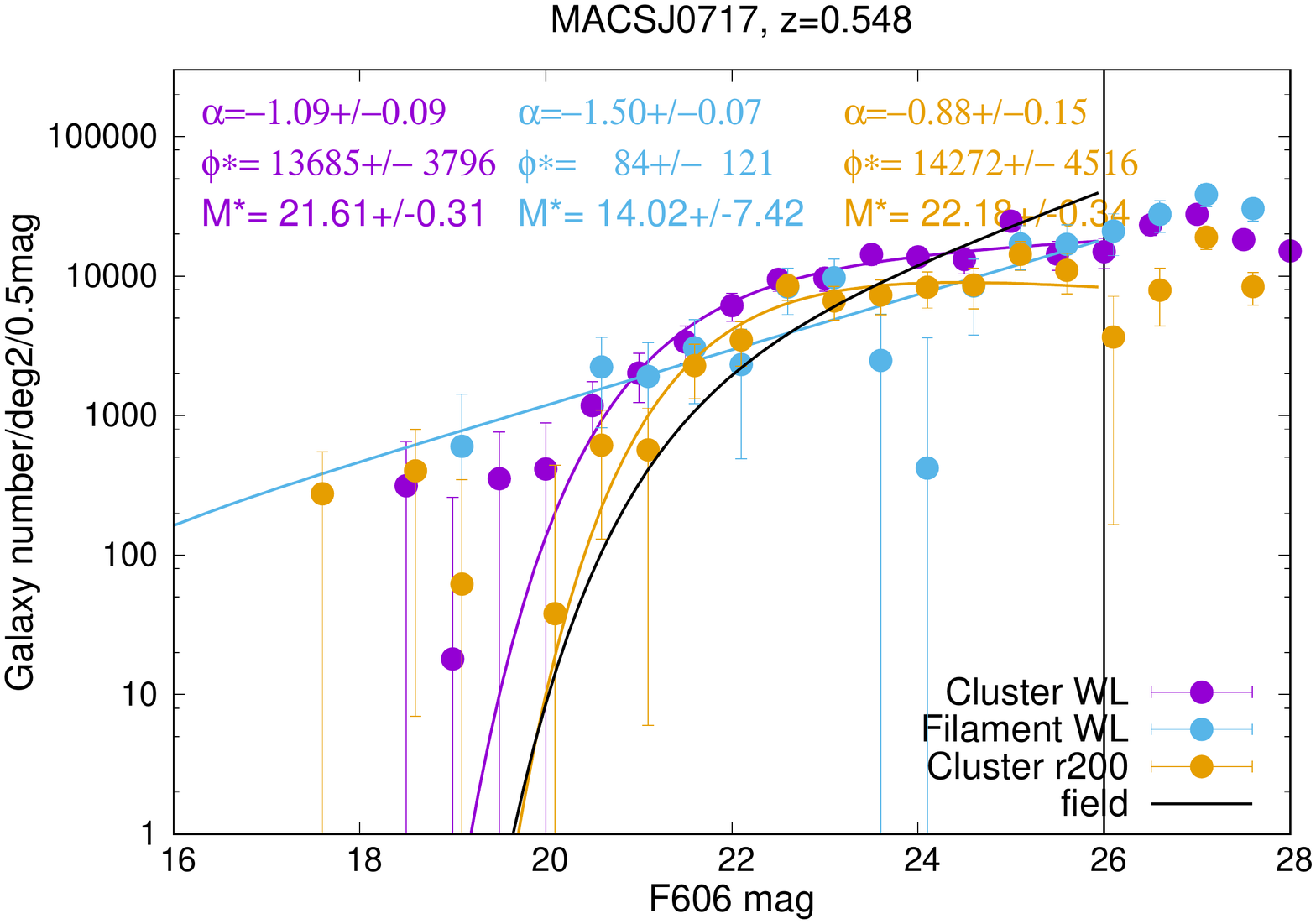}\\
  \includegraphics[width=3.5in,clip=true]{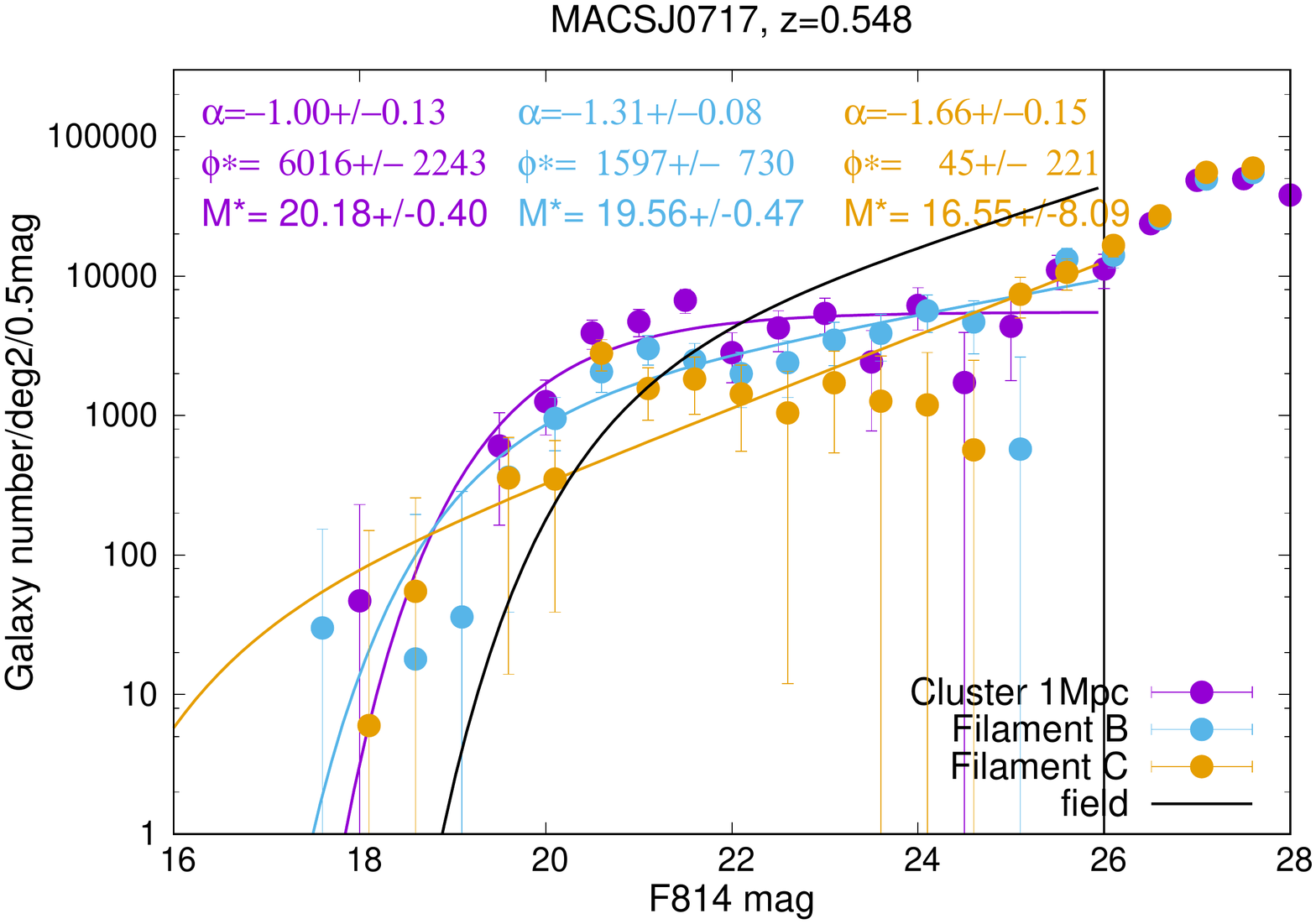}&\includegraphics[width=3.5in,clip=true]{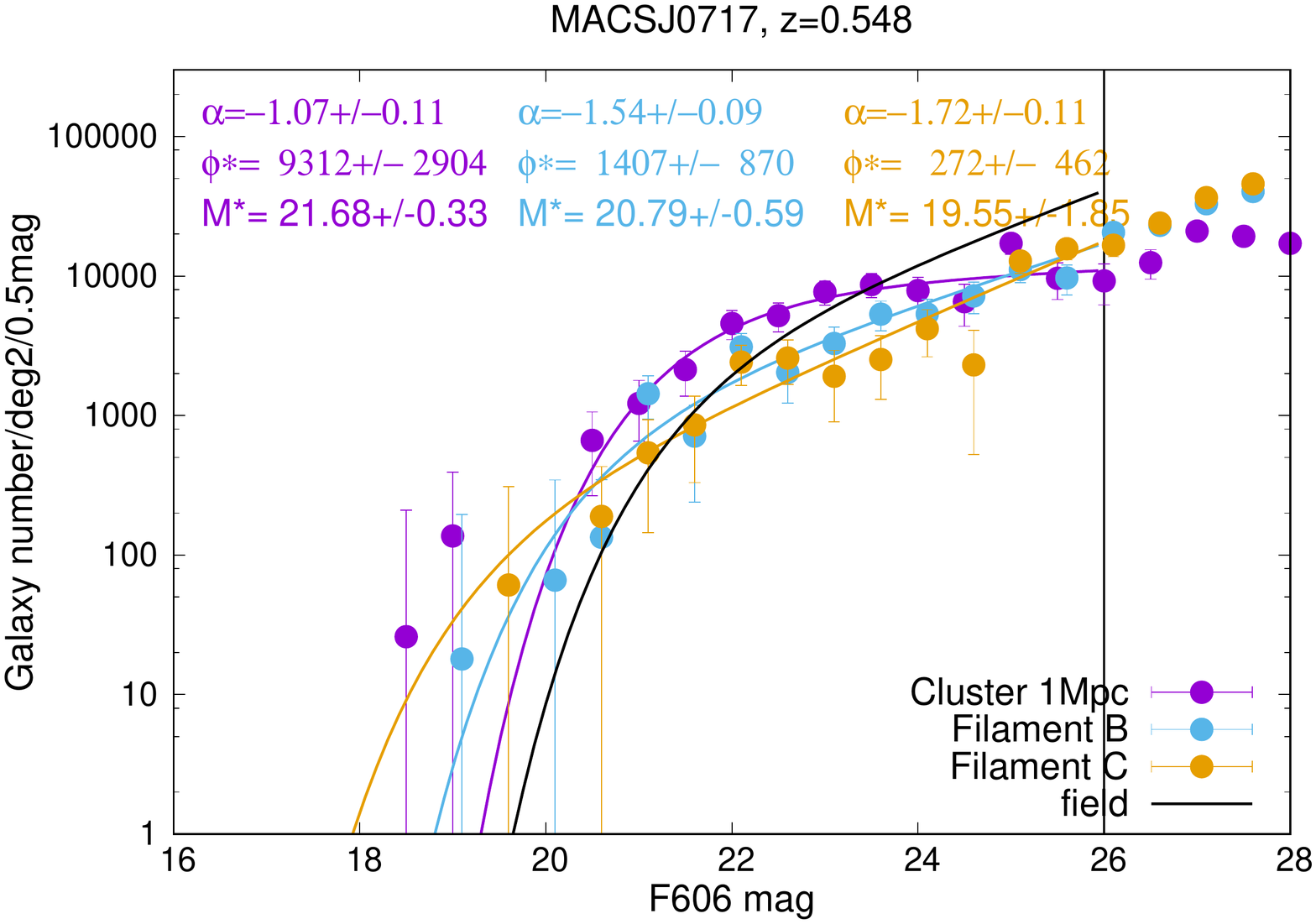}\\
  \end{tabular}
  \caption{Same as Fig.~\ref{fig:glf1} but for all galaxies (i.e. not only RS galaxies) in the F814W (\textit{left}) and F606 (\textit{right}) apparent magnitudes.}
  \label{fig:glftemp3}
\end{figure*}

Since we know that the RS may not be a good way of selecting filament galaxies, e.g. because they could correspond to field galaxies which should not lie at the cluster redshift, we also compute GLFs for all galaxies in apparent magnitudes, in F814W and F606W. The results (see Fig.~\ref{fig:glftemp3}) are very similar to what we find with the RS GLFs but with larger error bars due to a larger dilution by field galaxies. The disappearance of bright galaxies from F814W to F606W remains for filament B, and filament C shows no significant overdensity of bright galaxies in any band compared to the field.

What comes out of this study is that filament B shows a GLF between that of a cluster and that of the field, with an embedded pre-processed group detected in WL, and that filament C is a much poorer structure that might rather correspond to field galaxies. Finally, the difference between the F814W and F606W filters (especially in the case with all galaxies) tends to show that the galaxies in filaments close to clusters are preferentially passive than star forming.

\subsection{Selecting blue galaxies at the cluster redshift}
\label{sec:BS}

Without spectroscopic or photometric redshift information for all the galaxies, selecting objects inside a cosmological filament is a difficult task. In our case, having two magnitude bands allows for example to select red galaxies if we assume that galaxy populations in filaments are already preprocessed and show a red sequence. This assumption is probably true within the cluster-infalling groups embedded in the filaments, but more questionable for early-type galaxies that are not group members, a population which is probably not dominant in the filaments.

If we now consider late type galaxies, trying to select them in a filament by using the cluster blue cloud characteristics is probably impossible and unjustified. In an attempt to apply another way to select such blue type galaxies in the filament detected in the MACS~J0717 field of view, we chose to estimate the spatial distribution of these galaxies. Our present assumption is that their spatial distribution is not too different from that of red galaxies. Filaments of galaxies are structures of low mass and concentration, without a very
dense intra-cluster medium (X-ray emission of filaments is weak), and their potential well should therefore not affect very differently their red and blue galaxies as a function of their mass. This hypothesis is also supported by the studies of the two point correlation function of field galaxies. At similar redshifts, \citet{DeLaTorre2011} (see their Figure~12) show that blue and red galaxies have very similar two point correlation signals for correlation lengths lower than 0.2 Mpc and larger than 0.8 Mpc (to be compared with the typical size of the filaments studied here: $\sim$2$\times$4 Mpc).

We therefore chose to use the Minimal Spanning Tree technique (MST hereafter). This tool allows to characterize spatial distributions of points by tracing the tree of minimal length linking all the considered points \citep{Adami1999}. For a given set of points, the tree of minimal length is not unique, but the histogram of the lengths of its branches is unique. This histogram therefore fully characterizes a given distribution of points.

In our case, we considered galaxy distributions on the sky (a 2D approach) because we have no redshift information. As shown in \citet{Adami1999}, the optimal set of statistic descriptors of the histograms of branch lengths is the mean, $\mu$, the rms, $\sigma$, and the skewness, $\mathcal{S}$, of the branch lengths.

The general goal of our attempt is to find the non-red sequence galaxy populations having a 2D spatial distribution as close as possible to that of the RS galaxy population inside the area of the filament candidates. 

\begin{enumerate}

    \item We select all galaxies present on the line of sight of filaments B and C (see Fig.~\ref{fig:HSTmosaic}). 
    
    \item We select galaxies in the RS for these two areas, and compute $\mu$, $\sigma$, and $\mathcal{S}$ for the branch lengths of their MSTs. These RS galaxies are therefore supposed to be filament members. 
    
    \item We select in the same areas all galaxies outside of the red sequence (ORS hereafter). These samples include blue filament member galaxies plus foreground and background galaxies. If our assumption is true, the MST built on the blue filament member galaxies should have the same $\mu$, $\sigma$, and $\mathcal{S}$ as the MST of RS galaxies.
    
    \item The goal is then to find within each filament candidate the ORS galaxy sub-population being the closest to the RS galaxy population. This is technically done by computing a quadratic distance between the two populations in parameter space \citep{Adami1999}:
    $$
    D=\sqrt{(\mu_{\rm{RS}}-\mu_{\rm{ORS}})^{2}+(\sigma_{\rm{RS}}-\sigma_{\rm{ORS}})^{2}+(\mathcal{S}_{\rm{RS}}-\mathcal{S}_{\rm{ORS}})^2}
    $$

    \item We start with the total ORS galaxy sample and we search for a single galaxy to remove from the sample in order to have the largest diminution of $D$. This galaxy is supposed to be a foreground or background object, and is removed from the running ORS galaxy sample.

    \item The previous step is repeated iteratively. This allows us to draw Fig.~\ref{fig:msigske}, where we show the value of $D$ as a function of the number of removed galaxies. As expected, the curves show a minimum value of $D$, corresponding to the ORS galaxy population which is the most similar to the RS galaxy population in terms of spatial distribution.
    
    \item However, there are some uncertainties in our calculations, due to the intrinsic statistical errors in the estimations of $\mu$, $\sigma$, and $\mathcal{S}$. This results in a typical error bar shown as the horizontal line in Fig.~\ref{fig:msigske} (see \citet{Adami1999} for the estimate).
    
    \item We can then finally use Fig.~\ref{fig:msigske} to define three sub-samples within the ORS galaxy populations.
    
    \begin{itemize}
        \item The maximal sample (MAX hereafter): largest possible ORS galaxy sample with a $D$ value lower than the typical uncertainty.
        \item The minimal sample (MIN hereafter): smallest possible ORS galaxy sample with a $D$ value lower than the typical uncertainty.
        \item The optimal sample (OPT hereafter): ORS galaxy sample with the lowest possible $D$ value. 
    \end{itemize}
    
\end{enumerate}

\subsection{GLFs of blue galaxies}

\begin{figure}[!ht]
  \begin{center}
    \includegraphics[angle=270,width=3.5in,clip=true]{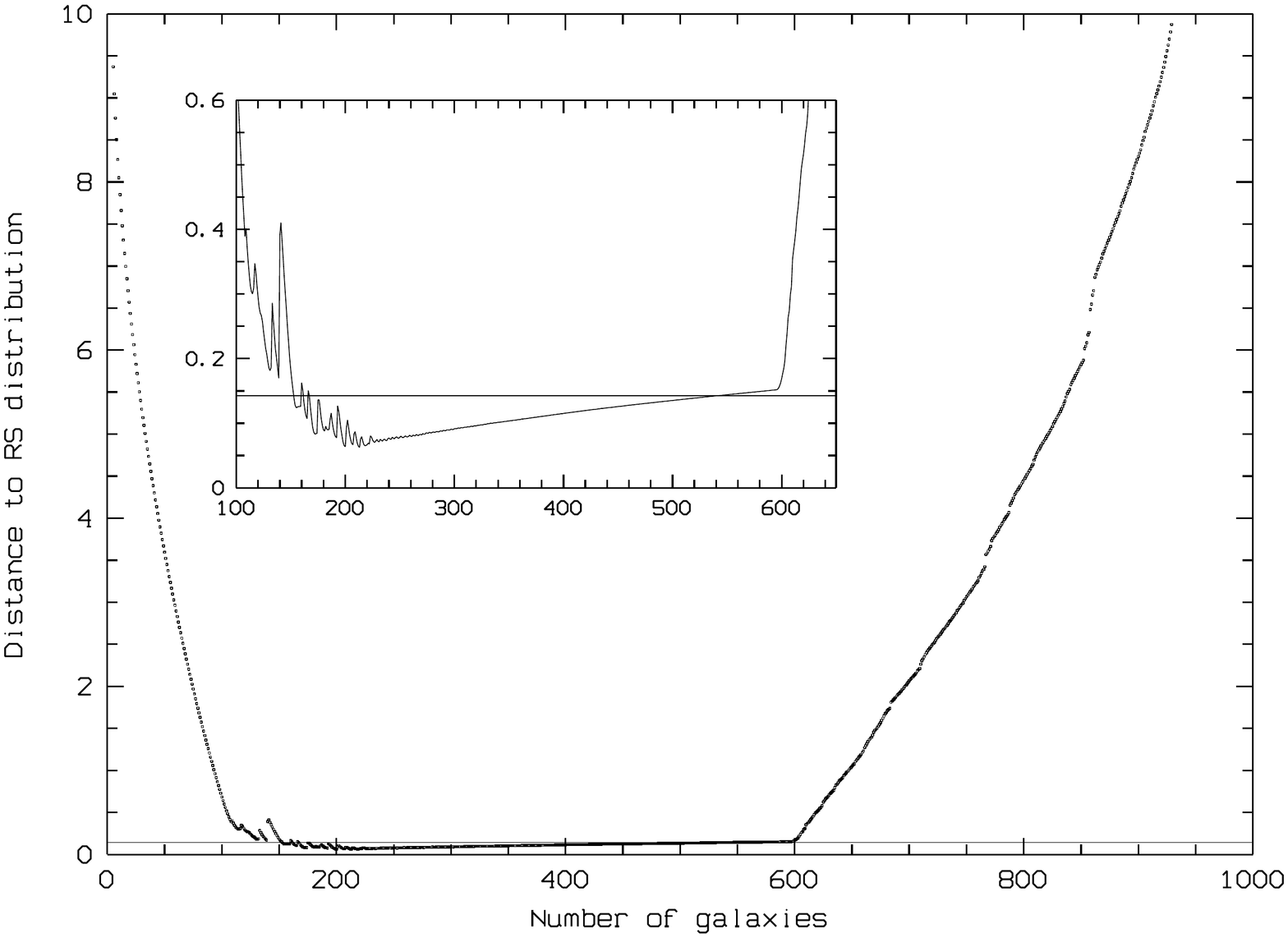}
    \includegraphics[angle=270,width=3.5in,clip=true]{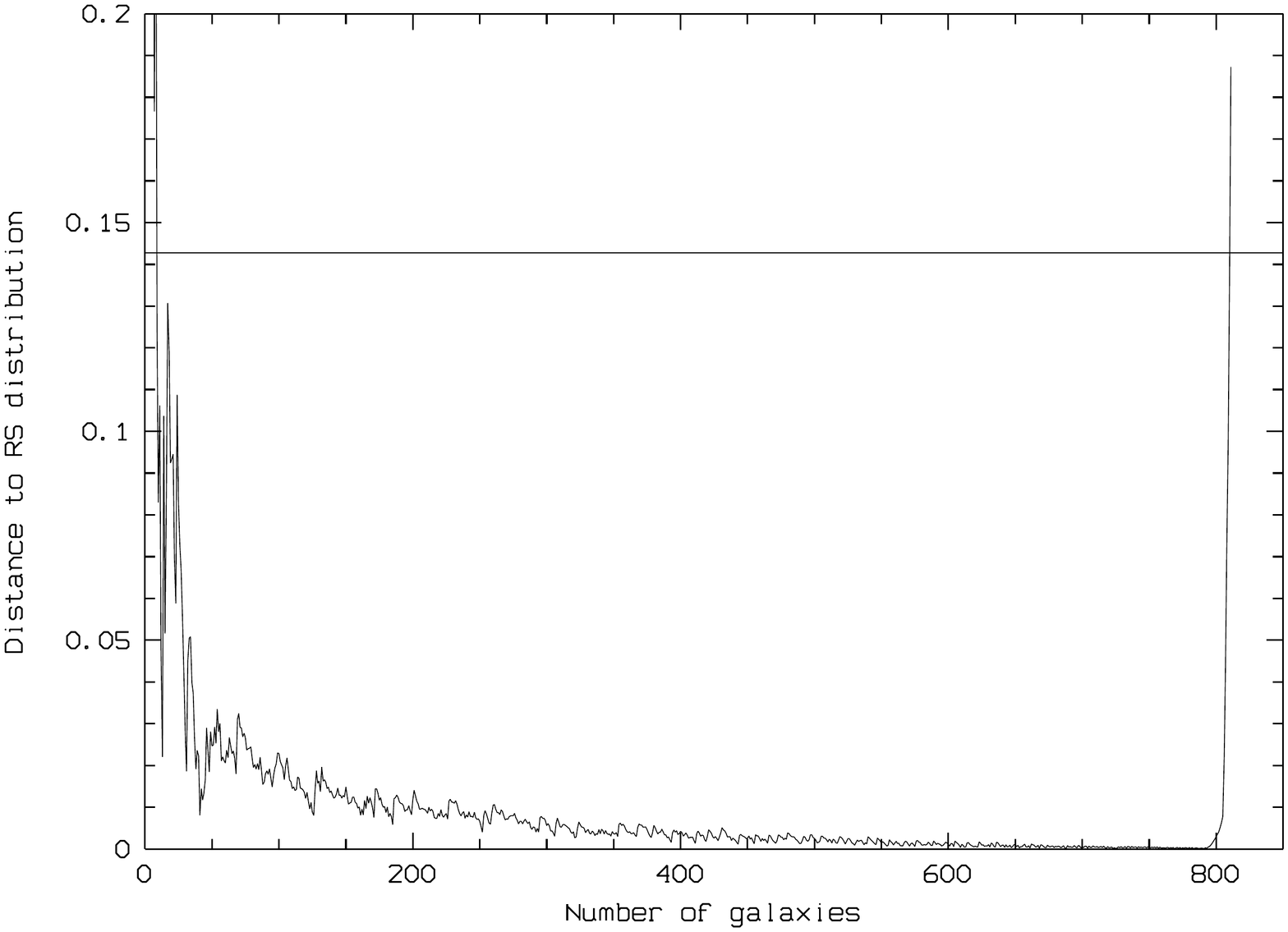}
    \caption{Distance D between RS and ORS galaxy samples versus number of galaxies in the ORS galaxy sample. The horizontal line is the typical uncertainty in the D measurement from \citet{Adami1999}. Top: filament B, bottom: filament C, with the inner plot showing a zoom on the region of interest.}
  \label{fig:msigske}
  \end{center}
\end{figure}

Figure~\ref{fig:msigske} clearly shows that for filament C, the ORS galaxy sample is very similar in terms of spatial distribution to the RS galaxy sample, whatever the selection within the ORS sample (D is nearly always lower than the typical uncertainty). This could mean that defining a red sequence in filament C is meaningless.

Filament B shows a clearer tendency of the ORS sample to be similar to the RS sample only between 150 and 600 galaxies. As compared to the initial $\sim$930 galaxies along the line of sight within the ORS sample, this means that we need to remove at least $\sim$35$\%$ of the galaxies along the line of sight to have similar spatial distributions between ORS and RS samples.

We analyzed in the same way the filament of galaxies detected between the A222 and A223 galaxy clusters \citep{Durret2010}. Despite being only poorly sampled
with spectroscopy (only five galaxy spectroscopic redshifts in the ORS sample: four within the filament and one outside), we find that the galaxies outside of the filament are the first to be removed in the process (well before D reaches its minimal value) while the galaxies inside the filament start to be removed just before D reaches its minimal value.

\begin{figure}[!ht]
  \begin{center}
    \includegraphics[angle=0,width=3.5in,clip=true]{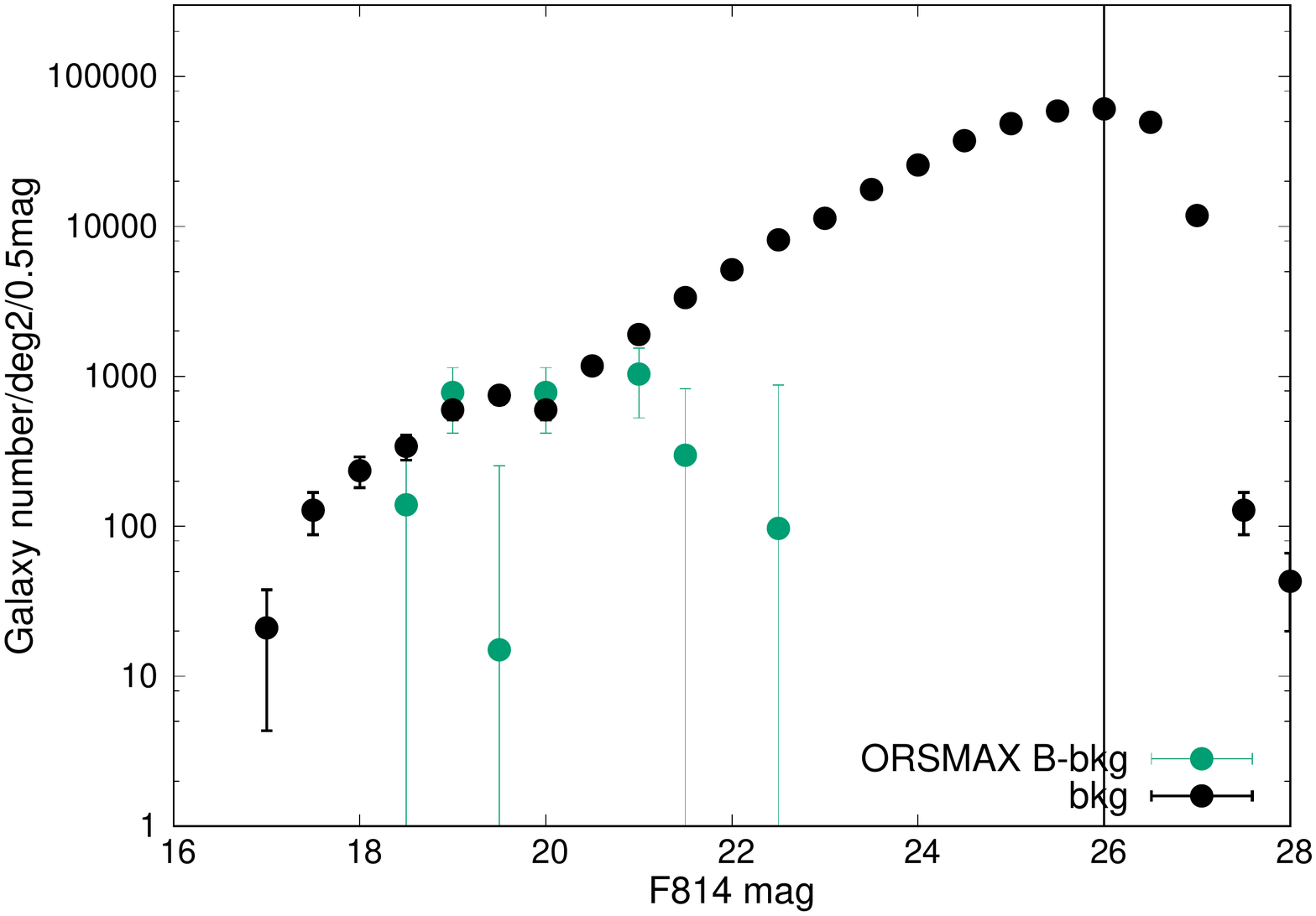}
    \includegraphics[angle=0,width=3.5in,clip=true]{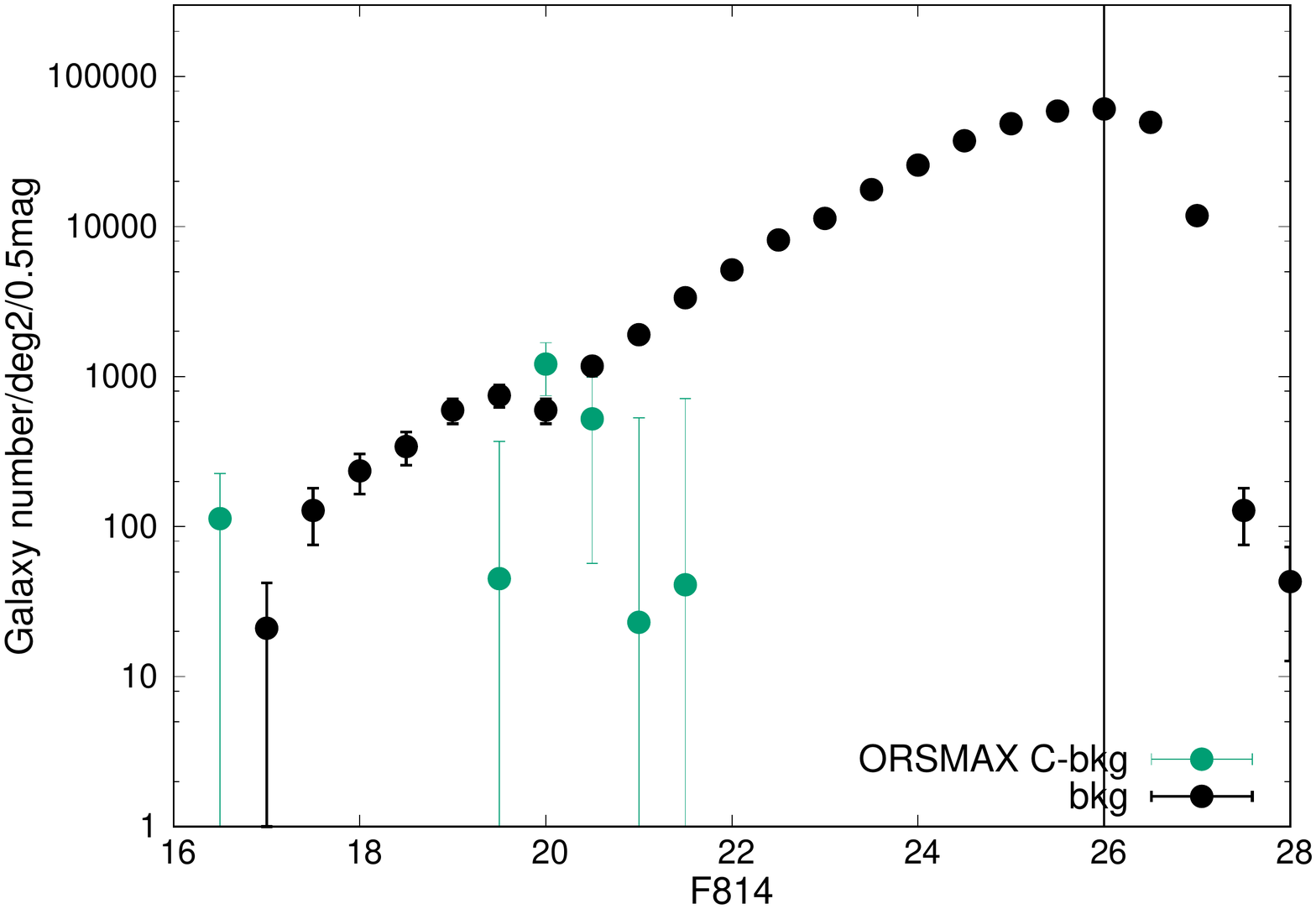}
    \caption{F814W apparent magnitude luminosity functions of samples of blue galaxies in filaments B (top figure) and C (bottom figure). Black disks represent the field, and green disks the filament blue samples after field subtraction.}
    \label{fig:BS_GLF}
  \end{center}
\end{figure}

We therefore decided to assume the ORS MAX sample as the galaxy sample representing in the best possible way the filament blue galaxies. With this sample, we compute the resulting galaxy luminosity function, following the method described in Sect.~\ref{subsec:compuGLF} (see Fig.~\ref{fig:BS_GLF}). Since we do not consider RS galaxies, it is not possible to compute an accurate k-correction in this case. In addition this approach requires to treat background galaxies differently. We apply a color cut to select field galaxies that lie outside of the cluster RS, and we also re-weight field galaxy counts by the ratio of the number of galaxies in the ORS catalog to the number of galaxies in the ORS MAX catalog to account for the dilution in the blue galaxy selection process. 

We derived the GLFs of these blue galaxies in the F606W and F814W bands for filaments B and C. The results are quite noisy, but one interesting feature is an excess of bright blue galaxies in filament B, with about twice as many blue galaxies brighter than M* than in the field. This suggests that a large group or small cluster, rich in bright blue galaxies, resides in region B and is merging with the main cluster, MACS~J0717. This group would be demarcated by the cyan contour in Fig.~\ref{fig:HSTmosaic}.

\section{Galaxy alignments}
\label{sec:PAs}

To gain more insight on the filament embedding MACS~J0717.5+3745, this section aims at assessing whether galaxies located in this structure show any preferential orientation of their major axis, or instead, if such orientations are random. As introduced in Section \ref{sec:intro}, this is motivated by previous findings of preferential directions for the orientations of filament galaxies - especially for some galaxy types or classes - relatively to the orientation of the filament ridge, be it in observations at low$-z$ \citep[e.g.][and references therein]{Tempel2013a,Tempel2013b,Zhang2015,Chen2019}, or in N-body and hydrodynamic simulations \citep[e.g.][and references therein]{Chen2015,Codis2015,Ganesh2018}. In both types of work, the underlying physical motivation for such alignments lies in tidal torque theory and mergers, as amply discussed in all these references. The assessment of this behavior is expected to put some constraints on galaxy and structure formation theories, and alignments have also been used to develop alternative methods of detecting filaments in the cosmic web around clusters \citep{Rong2016}. However, up to now, disparate results have been found, and the picture is far from clear. For instance, in contrast with previous results for the local Universe based on SDSS data, the recent work by \citet{Krolewski2019} uses MaNGA kinematic maps and finds no evidence for alignment between galaxy spins and filament directions. Our study on the filaments connected with MACS~J0717.5+3745 intends to add up to the observational determinations of galaxy orientations in filaments, this time in a structure at $z\sim0.5$.

Since filament B seems to be dominated by a group of galaxies (possibly located at the infall region of the cluster), as already discussed in the previous sections, we will restrict our analysis to galaxy orientations in filament C. We thus selected the frames from the \emph{HST} mosaic covering this region only (the green ellipse in Fig.~\ref{fig:HSTmosaic}). We note that filament C is almost totally covered by these observations that leave out only a small percentage (7.9\%) of its southern edge.

Because astrometry is accurate and homogeneous in this region, we can now adopt a slightly different approach from Sect.~\ref{sec:GLF}: as before, detections were made individually on each image of the \emph{HST} mosaic, but we now run \texttt{SExtractor} in double mode so as to directly obtain colors for the galaxies. 
We first ran this software on the deeper, less noisy, F814W images to detect sources, and compute their magnitude and peak surface brightness. Detection, background, deblending and aperture parameters were optimized especially for all objects that clearly stand out from the background, without a concern for achieving a thorough detection of faint galaxies that could be confused with noise or lost in brighter background areas. In particular, we eyeballed the apertures for photometry in the images to make sure that their extent was sufficient to fully cover galaxies down to the sky level and that their orientation looked correct. The hand-made masks covering all problematic areas (such as saturated stars and their bright spikes, as well as image edges) were used to discard spurious and contaminated detections at this stage and in all the subsequent analysis. 

For each image, we next selected stars in a magnitude versus maximum surface brightness plane (MU\_MAX versus MAG\_AUTO), and used the catalog of these objects as input for a second run of \texttt{SExtractor} to obtain the necessary files for \texttt{PSFEx} \citep{Bertin2011}, previously verifying that these stars were well spread all over the image. The PSF model computed with \texttt{PSFEx} for each image was saved, and fed into \texttt{SExtractor} in a third run for all images, now carried out in the ASSOC mode, choosing a de Vaucouleurs plus a disk model to fit all sources. This allowed us to compute PSF corrected magnitudes (MAG\_AUTO, which include the correction for Galactic extinction according to \citet{Schlafly2011}), position angles and ellipticities for all objects in all images in the F814W filter.
Finally, to compute galaxy colors, we did a final run of \texttt{SExtractor} in double mode, and using F814W as the detection images while magnitudes were now measured in the corresponding F606W images (adapting all necessary parameters such as magnitude zero point, extinction correction and gain). 

Catalogs obtained for all frames in the two filters were concatenated, and double entries were excluded, based on the criteria described in the previous section, for objects located where adjacent frames overlap. Finally, stars were purged from the catalog (based on their position in the MAG\_AUTO -- MU\_MAX plane), which was further limited to include only objects brighter than 25 mag in F814W. This last point minimizes the number of objects with large errors in the determination of the position angle.

At this stage, we selected galaxies with positions (RA, DEC) within the green ellipse of Fig.~\ref{fig:HSTmosaic} \citep[i.e. the 3$\sigma$ density contours that delineate filament C as defined in][]{Durret2016}.

The next step would be to identify all galaxies within this filament along the line-of-sight. As before, and in the absence of abundant spectroscopic coverage or photometric redshifts for this area, we will use the cluster red sequence to select galaxies in the filament, assuming through this colour indication that they lie at the same distance as cluster galaxies. This produces a sample of 390 galaxies. Among these, there will still be a percentage of field galaxies (up to $\sim 59\%$, as estimated from the field counts - Sect.~\ref{sec:GLF}), but since we don’t expect these objects to have a particular orientation, their presence will not affect the results.

\begin{figure}
  \centering
  \includegraphics[width=3.5in,clip=true]{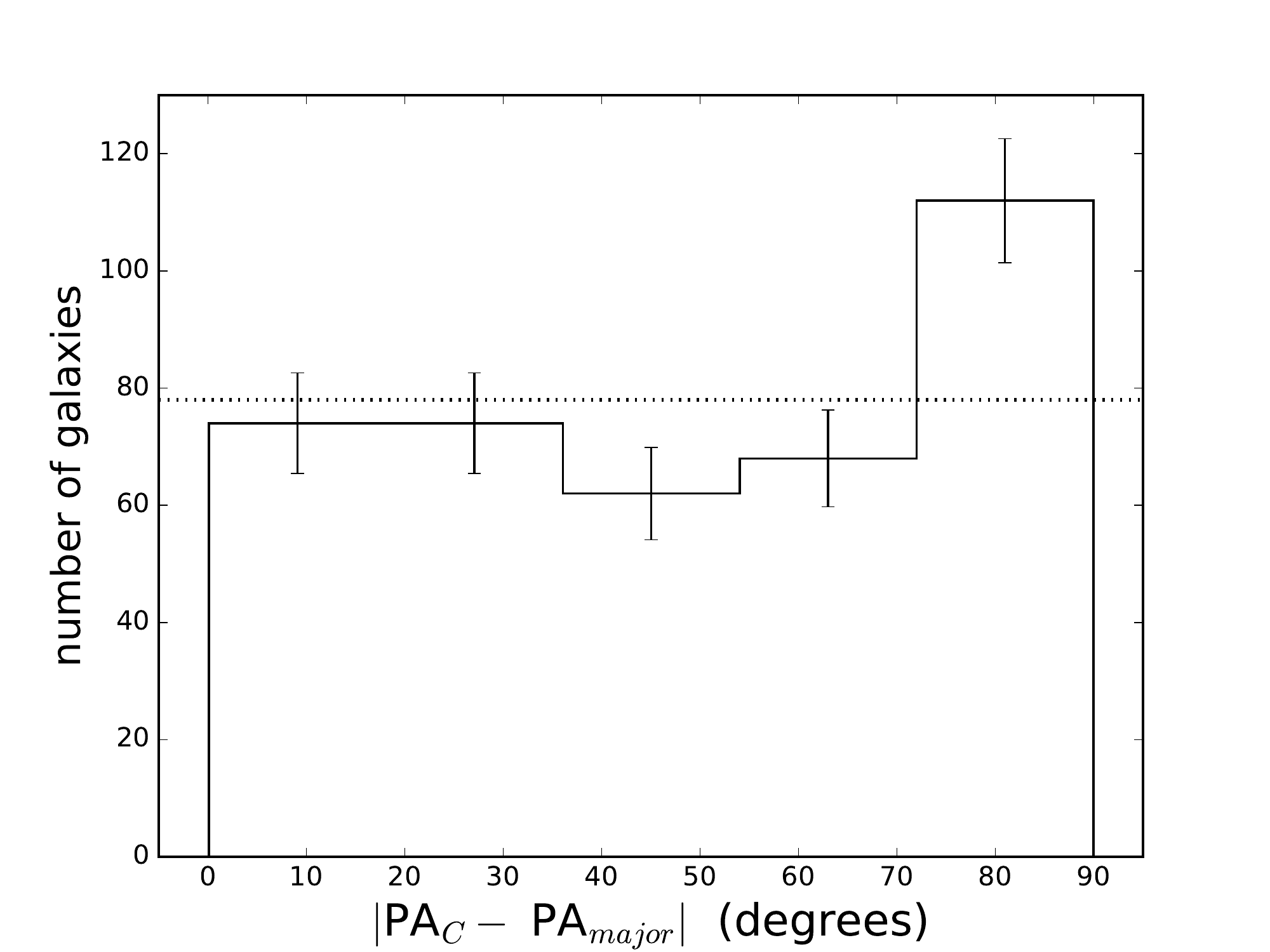}
  \caption{Distribution of the angles (in absolute value) subtended by the orientation of filament C and the orientation of the position angles of the 390 red sequence galaxies located in the same filament (within the 3$\sigma$ density contours - see text). Error bars assume Poissonian statistics and the dotted line represents the average number of galaxies per bin.}
  \label{fig:PAC}
\end{figure}

\begin{figure}
  \centering
  \includegraphics[width=3.5in,clip=true]{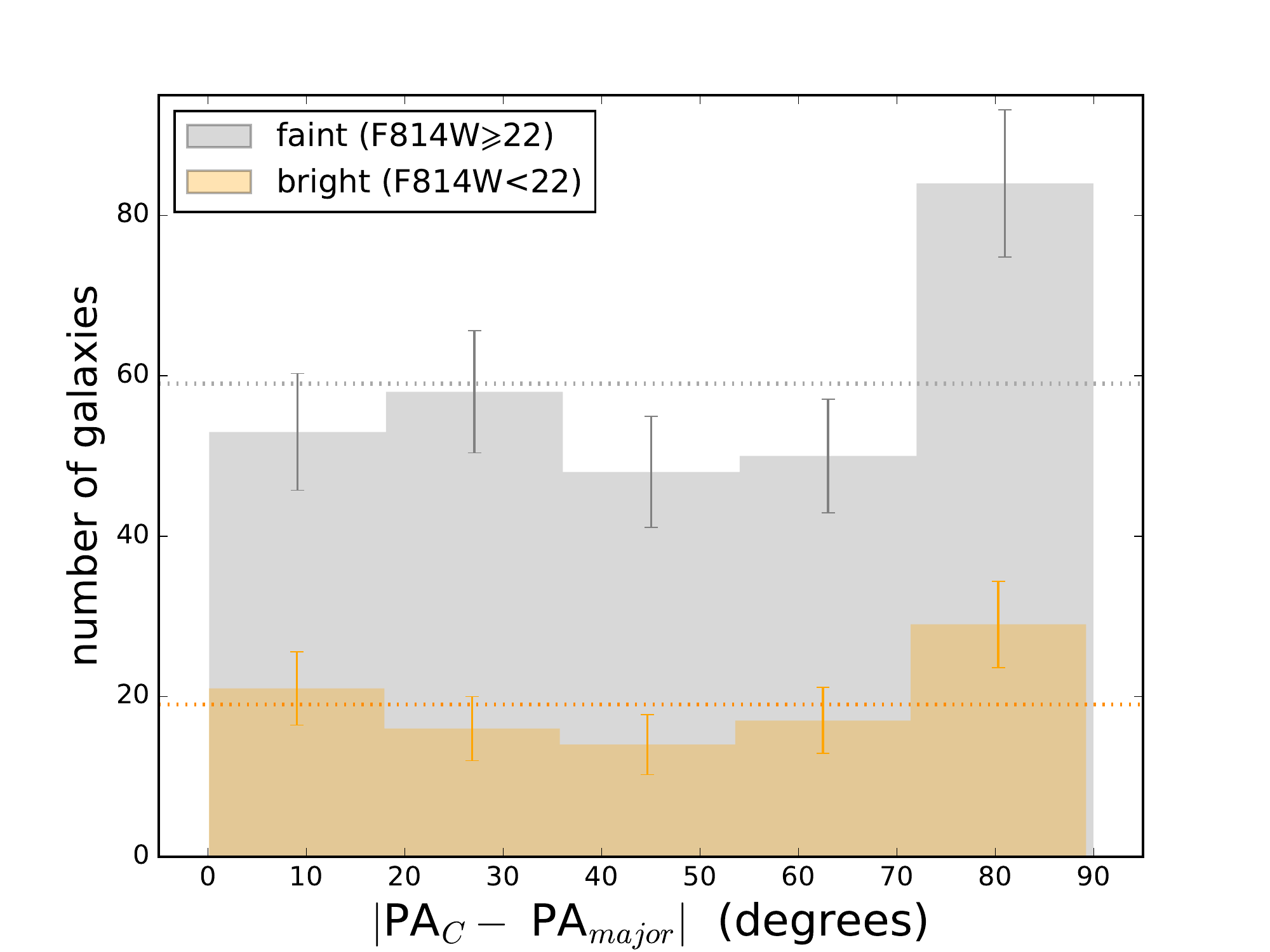}
  \caption{Same as Fig.~\ref{fig:PAC} but splitting the sample according to apparent magnitude in the F814W band: there are 97 bright galaxies and 293 faint galaxies.}
  \label{fig:PAC_mag}
\end{figure}

\begin{figure}
  \centering
  \includegraphics[width=3.5in,clip=true]{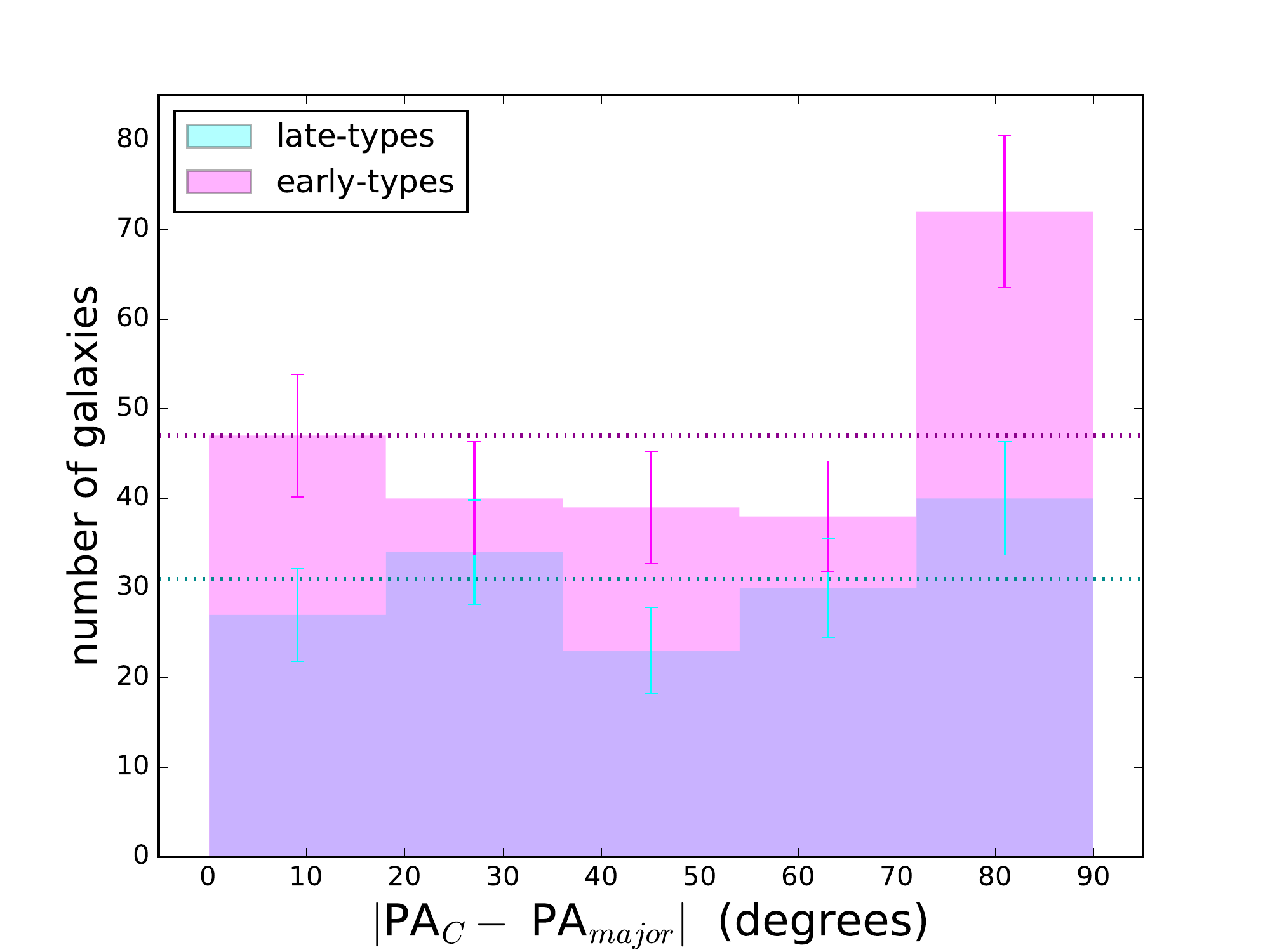}
  \caption{Same as Fig.~\ref{fig:PAC} but splitting the sample by galaxy type: 236 early-types against 154 late-types.}
  \label{fig:PAC_type}
\end{figure}

Figure~\ref{fig:PAC} shows the absolute value of the angle between the direction of filament C (which is oriented North-South) and the position angle of the major axis for all 390 galaxies included in the previously defined sample. 
There seems to be an excess of red sequence galaxies with their major axis oriented perpendicularly to the direction of the filament, although this is a low significance trend (a departure of less than $3\sigma$ from the mean behavior). The Anderson-Darling test issues a significance level of 4\% when  comparing the two distributions of Fig.~\ref{fig:PAC}. According to the classical interpretation of the p-value in such statistical tests, a significance level of 4\% is usually considered to be sufficiently low to conclude that the null hypothesis (i.e. the two distributions are alike) is clearly disfavored. However, this value lies at the borderline  (usually taken to be 5\%) between accepting or rejecting the null hypothesis; if we took into account the error bars, then the expected significance level would likely increase, rendering the test result inconclusive. 

To understand, however, whether the trend seen in Figure~\ref{fig:PAC} is caused by a specific class of RS galaxies, we split the sample by type and brightness. As a proxy for the morphological type, we consider the bulge-to-total flux ratio in the F814W band, setting the frontier at 0.35 for this quantity as in \citet{Simard2009}. As for brightness, we set the separation at 22\,mag in F814W, so as to keep a reasonable number of objects in the bright set (different alternative limits were tested without relevant changes in the outcome). For this new brightness cut, the percentage of field galaxies possibly contaminating the sample lowers to about $46\%$. Figures~\ref{fig:PAC_mag} and  \ref{fig:PAC_type} show the results after separating the sample according to each of these two criteria.

Figure~\ref{fig:PAC_mag} shows it is faint galaxies (which are also dominant in number) that are preferentially oriented perpendicularly to the filament, and these are likely of early-types (Figure~\ref{fig:PAC_type}). We find no clear indication for a preferential alignment for either bright or late-types, that are much less numerous. However, we note that a sturdy analysis based on morphologies is hampered both by our crude classification and by the known difficulty of determining position angles for pure ellipticals (that will make up a certain percentage of our early-types) using photometric data only. 

It is difficult to compare our result with other observational determinations, and also with the published results based on simulations, due to our selection: restricting our analysis to red sequence galaxies limits our capacity to infer definite conclusions. Still, our study of region C provides a low-significance indication of an excess of faint galaxies having their major axis aligned along a direction perpendicular to the filament. 

As a robustness test of this result, we repeated the same analysis by restricting our sample to within the 4$\sigma$ density contour region defined by \citet{Durret2016}, without any relevant change in the results. We further consider that any errors in the determination of the position angle, likely affecting more severely fainter galaxies, should not be responsible for the appearance of a specific trend.
Finally, if we abandon our selection criteria, and use all available spectroscopic data (as in NED) for filament C instead, we are left with 61 galaxies having a redshift compatible with the cluster redshift range, i.e. [0.530,0.560] \citep{Durret2016} – this is too small a sample for our objectives, further plagued with incompleteness in terms of limiting magnitude for the spectroscopic coverage. Results are thus noisier and quite inconclusive. 

In summary: our results for region C point to there being no preferential orientation of bright RS galaxies, whereas a preferential orientation of faint RS galaxies might exist, in which the major axes of galaxies lie mainly perpendicular to the direction of the filament. Such a result is of low significance though (below $3\sigma$ from random alignment) but, if we consider it, this seems to be in contrast with what authors have been finding at lower redshifts based on SDSS data, where small but significant alignment in the direction parallel to the orientation of nearby filaments where found \citep[][and other works of these teams]{Tempel2013a,Tempel2013b,Zhang2015,Chen2019}. 

Three reasons could be proposed to explain this discrepancy: (i) there is no filament in region C (as hinted by the results of Sect.~\ref{sec:GLF}); (ii) the filament does exist \citep[as inferred by][ even if it is not a very dense structure given its low significance detection in weak lensing and absence of X-ray emission]{Durret2016}, but we are unable to correctly identify its member galaxies; (iii) the filament does exist, and we are correctly sampling its population (even if only a subset of it), and the fact that we do not obtain the trends in orientation that are found in low-z systems needs to be understood. 
This could be a real result, explained by galaxy evolution within the filaments or simply by the intrinsic variety of filament properties, or something induced by errors - either in our assessment of the position angles or in the estimates provided by other authors, possibly worsened by the lower quality SDSS images. Interestingly enough, in a search for possible radial alignments in massive clusters located at z>0.5, and observed with \emph{HST} (MACS~J0717.5+3745 included), \citet{Hung2012} found no such alignments, in stark contrast to what is reported for nearby clusters at z$\sim$0.1 using SDSS data. These authors discuss possible explanations for the discrepancy, invoking either evolutionary effects or the {\em ``presence of systematic biases in
the analysis of SDSS imaging data that cause at least partly spurious alignment signals.''} It is thus plausible that similar causes could explain the lack of agreement between our \emph{HST}-based results for a z$\sim$0.5 filament and trends obtained for lower redshift filaments in SDSS data. Finally, 

Future determinations of galaxy orientations in this filament, carried out with the aid of spectroscopic data, to select only galaxies at the cluster redshift, and in other filaments around z=0.5 will be of great help in disentangling this issue.

\section{Conclusions}
\label{sec:conclusions}

We analyzed the properties of MACS~J0717's cosmic filament by computing the GLF in different regions (B and C). While the RS GLF of region C is that of a poor structure close to that of field galaxies, the RS GLF of region B is much richer and typical of a galaxy group. This is consistent with the presence of WL contours in area B corresponding to a denser area. Recent work has also detected this structure and estimated its mass within 150~kpc \citep[$ M_{150}=(2.28 \pm 0.24)\times 10^{13}M_{\odot}$,][]{Jauzac2018}. When looking at the GLFs of blue galaxies, an excess of bright blue galaxies in region B leads to the same conclusion of a rich galaxy group moving along the cosmic filament to merge with MACS~J0717. Merging phenomena in the filament could therefore take place preferentially for massive blue galaxies, with some gas remaining, allowing them to stay blue.

We also studied the orientation of RS galaxies in filament C, finding no preferential orientation of bright RS galaxies. A preferential orientation perpendicular to the direction of the filament for faint galaxies might exist but is below the 3$\sigma$ level of significance from a random distribution, making it difficult to draw conclusions.

We used our new software \texttt{DAWIS} to detect and estimate the ICL of the core of MACS~J0717 in the HFF, finding ICL fractions coherent with the literature. The fact that we detect almost no ICL in the UV rest frame of MACS~J0717 indicates that there must be little star formation in the ICL. This is also confirmed by the fact that the ICL in the F105W band (that contains the H$\alpha$ line) is also rather weak. This result agrees with results based on integral field spectroscopy with VLT/MUSE by \cite{Adami2016} for a cluster (XLSSC~116) at a similar redshift (see their Table~2).

We tried to detect large diffuse sources in the filament that could be associated with IFL, without success. We found a single system of strongly disturbed galaxies with obvious tidal streams lying inside the WL contours of the galaxy group falling into MACS~J0717. This supports the ICL formation scenario in which such galaxy groups are forming IGL through tidal stripping and mergers, and then fall into bigger structures, mixing their IGL into the bigger structure's ICL.

\begin{acknowledgements}
We acknowledge long-term support from CNES. C.~Lobo thanks the IAP for inviting her for a week and further acknowledges support by Fundação para a Ciência e a Tecnologia (FCT) through national funds (UID/FIS/04434/2013) and by FEDER - Fundo Europeu de Desenvolvimento Regional through COMPETE2020 - Programa Operacional Competitividade e Internacionalização (POCI-01-0145-FEDER-007672), and by FCT/MCTES through national funds (PIDDAC) via grant UID/FIS/04434/2019. N.~Martinet acknowledges support from a CNES fellowship.
M.~Jauzac acknowledges support by the Science and Technology Facilities Council (grant number ST/L00075X/1).
This research has made use of the NASA/IPAC Extragalactic Database (NED), which is operated by the Jet Propulsion Laboratory, California Institute of Technology, under contract with the National Aeronautics and Space Administration. We acknowledge Mark Taylor for the development of TOPCAT \citep{Taylor2005}.

\end{acknowledgements}


\bibliographystyle{aa}  
\bibliography{aa} 

\begin{thebibliography}{121}
\expandafter\ifx\csname natexlab\endcsname\relax\def\natexlab#1{#1}\fi

\bibitem[{{Adami} {et~al.}(2013){Adami}, {Durret}, {Guennou}, \& {Da
  Rocha}}]{Adami2013}
{Adami}, C., {Durret}, F., {Guennou}, L., \& {Da Rocha}, C. 2013, \aap, 551,
  A20

\bibitem[{{Adami} \& {Mazure}(1999)}]{Adami1999}
{Adami}, C. \& {Mazure}, A. 1999, \aaps, 134, 393

\bibitem[{{Adami} {et~al.}(2016){Adami}, {Pompei}, {Sadibekova}, {Clerc},
  {Iovino}, {McGee}, {Guennou}, {Birkinshaw}, {Horellou}, {Maurogordato},
  {Pacaud}, {Pierre}, {Poggianti}, \& {Willis}}]{Adami2016}
{Adami}, C., {Pompei}, E., {Sadibekova}, T., {et~al.} 2016, \aap, 592, A7

\bibitem[{{Adami} {et~al.}(2005){Adami}, {Slezak}, {Durret}, {Conselice},
  {Cuillandre}, {Gallagher}, {Mazure}, {Pell{\'o}}, {Picat}, \&
  {Ulmer}}]{Adami2005}
{Adami}, C., {Slezak}, E., {Durret}, F., {et~al.} 2005, \aap, 429, 39

\bibitem[{{Aguerri} {et~al.}(2006){Aguerri}, {Castro-Rodr{\'{\i}}guez},
  {Napolitano}, {Arnaboldi}, \& {Gerhard}}]{Aguerri2006}
{Aguerri}, J.~A.~L., {Castro-Rodr{\'{\i}}guez}, N., {Napolitano}, N.,
  {Arnaboldi}, M., \& {Gerhard}, O. 2006, \aap, 457, 771

\bibitem[{{Andreon}(2006)}]{Andreon2006}
{Andreon}, S. 2006, \mnras, 369, 969

\bibitem[{{Arnouts} {et~al.}(1999){Arnouts}, {Cristiani}, {Moscardini},
  {Matarrese}, {Lucchin}, {Fontana}, \& {Giallongo}}]{Arnouts1999}
{Arnouts}, S., {Cristiani}, S., {Moscardini}, L., {et~al.} 1999, \mnras, 310,
  540

\bibitem[{{Bertin}(2011)}]{Bertin2011}
{Bertin}, E. 2011, in Astronomical Society of the Pacific Conference Series,
  Vol. 442, Astronomical Data Analysis Software and Systems XX, ed. I.~N.
  {Evans}, A.~{Accomazzi}, D.~J. {Mink}, \& A.~H. {Rots}, 435

\bibitem[{{Bertin}(2013)}]{Bertin2013}
{Bertin}, E. 2013, {PSFEx: Point Spread Function Extractor}, Astrophysics
  Source Code Library

\bibitem[{{Bertin} \& {Arnouts}(1996)}]{Bertin1996}
{Bertin}, E. \& {Arnouts}, S. 1996, \aaps, 117, 393

\bibitem[{Bijaoui \& Ru{\'{e}}(1995)}]{Bijaoui1995}
Bijaoui, A. \& Ru{\'{e}}, F. 1995, Signal Processing, 46, 345

\bibitem[{{Bond} {et~al.}(1996){Bond}, {Kofman}, \& {Pogosyan}}]{Bond1996}
{Bond}, J.~R., {Kofman}, L., \& {Pogosyan}, D. 1996, \nat, 380, 603

\bibitem[{{Brammer} {et~al.}(2012){Brammer}, {van Dokkum}, {Franx},
  {Fumagalli}, {Patel}, {Rix}, {Skelton}, {Kriek}, {Nelson}, {Schmidt},
  {Bezanson}, {da Cunha}, {Erb}, {Fan}, {F{\"o}rster Schreiber}, {Illingworth},
  {Labb{\'e}}, {Leja}, {Lundgren}, {Magee}, {Marchesini}, {McCarthy},
  {Momcheva}, {Muzzin}, {Quadri}, {Steidel}, {Tal}, {Wake}, {Whitaker}, \&
  {Williams}}]{Brammer2012}
{Brammer}, G.~B., {van Dokkum}, P.~G., {Franx}, M., {et~al.} 2012, \apjs, 200,
  13

\bibitem[{{Burke} {et~al.}(2012){Burke}, {Collins}, {Stott}, \&
  {Hilton}}]{Burke2012}
{Burke}, C., {Collins}, C.~A., {Stott}, J.~P., \& {Hilton}, M. 2012, \mnras,
  425, 2058

\bibitem[{{Byrd} \& {Valtonen}(1990)}]{Byrd1990}
{Byrd}, G. \& {Valtonen}, M. 1990, \apj, 350, 89

\bibitem[{{Capaccioli} \& {de Vaucouleurs}(1983)}]{Capaccioli1983}
{Capaccioli}, M. \& {de Vaucouleurs}, G. 1983, \apjs, 52, 465

\bibitem[{{Chen} {et~al.}(2019){Chen}, {Ho}, {Blazek}, {He}, {Mandelbaum},
  {Melchior}, \& {Singh}}]{Chen2019}
{Chen}, Y.-C., {Ho}, S., {Blazek}, J., {et~al.} 2019, \mnras, 485, 2492

\bibitem[{{Chen} {et~al.}(2017){Chen}, {Ho}, {Mandelbaum}, {Bahcall},
  {Brownstein}, {Freeman}, {Genovese}, {Schneider}, \& {Wasserman}}]{Chen2017}
{Chen}, Y.-C., {Ho}, S., {Mandelbaum}, R., {et~al.} 2017, \mnras, 466, 1880

\bibitem[{{Chen} {et~al.}(2015){Chen}, {Ho}, {Tenneti}, {Mandelbaum}, {Croft},
  {DiMatteo}, {Freeman}, {Genovese}, \& {Wasserman}}]{Chen2015}
{Chen}, Y.-C., {Ho}, S., {Tenneti}, A., {et~al.} 2015, \mnras, 454, 3341

\bibitem[{{Codis} {et~al.}(2015){Codis}, {Gavazzi}, {Dubois}, {Pichon},
  {Benabed}, {Desjacques}, {Pogosyan}, {Devriendt}, \& {Slyz}}]{Codis2015}
{Codis}, S., {Gavazzi}, R., {Dubois}, Y., {et~al.} 2015, \mnras, 448, 3391

\bibitem[{{Contini} {et~al.}(2014){Contini}, {De Lucia}, {Villalobos}, \&
  {Borgani}}]{Contini2014}
{Contini}, E., {De Lucia}, G., {Villalobos}, {\'A}., \& {Borgani}, S. 2014,
  \mnras, 437, 3787

\bibitem[{{Contini} {et~al.}(2018){Contini}, {Yi}, \& {Kang}}]{Contini2018}
{Contini}, E., {Yi}, S.~K., \& {Kang}, X. 2018, \mnras, 479, 932

\bibitem[{{Da Rocha} \& {Mendes de Oliveira}(2005)}]{DaRocha2005}
{Da Rocha}, C. \& {Mendes de Oliveira}, C. 2005, \mnras, 364, 1069

\bibitem[{{Da Rocha} {et~al.}(2008){Da Rocha}, {Ziegler}, \& {Mendes de
  Oliveira}}]{DaRocha2008}
{Da Rocha}, C., {Ziegler}, B.~L., \& {Mendes de Oliveira}, C. 2008, \mnras,
  388, 1433

\bibitem[{{de Jong}(2008)}]{deJong2008}
{de Jong}, R.~S. 2008, \mnras, 388, 1521

\bibitem[{{de la Torre} {et~al.}(2011){de la Torre}, {Meneux}, {De Lucia},
  {Blaizot}, {Le F{\`e}vre}, {Garilli}, {Cucciati}, {Mellier}, {Pollo},
  {Abbas}, {Bottini}, {Le Brun}, {Maccagni}, {Scodeggio}, {Tresse},
  {Vettolani}, {Zanichelli}, {Adami}, {Arnouts}, {Bardelli}, {Bolzonella},
  {Cappi}, {Charlot}, {Ciliegi}, {Contini}, {Foucaud}, {Franzetti},
  {Gavignaud}, {Guzzo}, {Ilbert}, {Iovino}, {McCracken}, {Marinoni}, {Mazure},
  {Merighi}, {Paltani}, {Pell{\'o}}, {Pozzetti}, {Vergani}, {Zamorani}, \&
  {Zucca}}]{DeLaTorre2011}
{de la Torre}, S., {Meneux}, B., {De Lucia}, G., {et~al.} 2011, \aap, 525, A125

\bibitem[{{De Lucia} {et~al.}(2004){De Lucia}, {Poggianti},
  {Arag{\'o}n-Salamanca}, {Clowe}, {Halliday}, {Jablonka}, {Milvang-Jensen},
  {Pell{\'o}}, {Poirier}, {Rudnick}, {Saglia}, {Simard}, \&
  {White}}]{DeLucia2004}
{De Lucia}, G., {Poggianti}, B.~M., {Arag{\'o}n-Salamanca}, A., {et~al.} 2004,
  \apjl, 610, L77

\bibitem[{{De Lucia} {et~al.}(2007){De Lucia}, {Poggianti},
  {Arag{\'o}n-Salamanca}, {White}, {Zaritsky}, {Clowe}, {Halliday}, {Jablonka},
  {von der Linden}, {Milvang-Jensen}, {Pell{\'o}}, {Rudnick}, {Saglia}, \&
  {Simard}}]{DeLucia2007}
{De Lucia}, G., {Poggianti}, B.~M., {Arag{\'o}n-Salamanca}, A., {et~al.} 2007,
  \mnras, 374, 809

\bibitem[{{De Propris} {et~al.}(2013){De Propris}, {Phillipps}, \&
  {Bremer}}]{DePropris2013}
{De Propris}, R., {Phillipps}, S., \& {Bremer}, M.~N. 2013, \mnras, 434, 3469

\bibitem[{{DeMaio} {et~al.}(2018){DeMaio}, {Gonzalez}, {Zabludoff}, {Zaritsky},
  {Connor}, {Donahue}, \& {Mulchaey}}]{DeMaio2018}
{DeMaio}, T., {Gonzalez}, A.~H., {Zabludoff}, A., {et~al.} 2018, \mnras, 474,
  3009

\bibitem[{{Diego} {et~al.}(2015){Diego}, {Broadhurst}, {Zitrin}, {Lam}, {Lim},
  {Ford}, \& {Zheng}}]{Diego2015}
{Diego}, J.~M., {Broadhurst}, T., {Zitrin}, A., {et~al.} 2015, \mnras, 451,
  3920

\bibitem[{{Dietrich} {et~al.}(2012){Dietrich}, {Werner}, {Clowe}, {Finoguenov},
  {Kitching}, {Miller}, \& {Simionescu}}]{Dietrich2012}
{Dietrich}, J.~P., {Werner}, N., {Clowe}, D., {et~al.} 2012, \nat, 487, 202

\bibitem[{{Dubois} {et~al.}(2014){Dubois}, {Pichon}, {Welker}, {Le Borgne},
  {Devriendt}, {Laigle}, {Codis}, {Pogosyan}, {Arnouts}, {Benabed}, {Bertin},
  {Blaizot}, {Bouchet}, {Cardoso}, {Colombi}, {de Lapparent}, {Desjacques},
  {Gavazzi}, {Kassin}, {Kimm}, {McCracken}, {Milliard}, {Peirani}, {Prunet},
  {Rouberol}, {Silk}, {Slyz}, {Sousbie}, {Teyssier}, {Tresse}, {Treyer},
  {Vibert}, \& {Volonteri}}]{Dubois2014}
{Dubois}, Y., {Pichon}, C., {Welker}, C., {et~al.} 2014, \mnras, 444, 1453

\bibitem[{{Durret} {et~al.}(2010){Durret}, {Lagan{\'a}}, {Adami}, \&
  {Bertin}}]{Durret2010}
{Durret}, F., {Lagan{\'a}}, T.~F., {Adami}, C., \& {Bertin}, E. 2010, \aap,
  517, A94

\bibitem[{{Durret} {et~al.}(2011){Durret}, {Lagan{\'a}}, \&
  {Haider}}]{Durret2011}
{Durret}, F., {Lagan{\'a}}, T.~F., \& {Haider}, M. 2011, \aap, 529, A38

\bibitem[{{Durret} {et~al.}(2016){Durret}, {M{\'a}rquez}, {Acebr{\'o}n},
  {Adami}, {Cabrera-Lavers}, {Capelato}, {Martinet}, {Sarron}, \&
  {Ulmer}}]{Durret2016}
{Durret}, F., {M{\'a}rquez}, I., {Acebr{\'o}n}, A., {et~al.} 2016, \aap, 588,
  A69

\bibitem[{{Ebeling} {et~al.}(2004){Ebeling}, {Barrett}, \&
  {Donovan}}]{Ebeling2004}
{Ebeling}, H., {Barrett}, E., \& {Donovan}, D. 2004, \apjl, 609, L49

\bibitem[{{Eckert} {et~al.}(2015){Eckert}, {Jauzac}, {Shan}, {Kneib}, {Erben},
  {Israel}, {Jullo}, {Klein}, {Massey}, {Richard}, \& {Tchernin}}]{Eckert2015}
{Eckert}, D., {Jauzac}, M., {Shan}, H., {et~al.} 2015, \nat, 528, 105

\bibitem[{{Fukugita} {et~al.}(1995){Fukugita}, {Shimasaku}, \&
  {Ichikawa}}]{Fukugita1995}
{Fukugita}, M., {Shimasaku}, K., \& {Ichikawa}, T. 1995, \pasp, 107, 945

\bibitem[{{Ganeshaiah Veena} {et~al.}(2018){Ganeshaiah Veena}, {Cautun}, {van
  de Weygaert}, {Tempel}, {Jones}, {Rieder}, \& {Frenk}}]{Ganesh2018}
{Ganeshaiah Veena}, P., {Cautun}, M., {van de Weygaert}, R., {et~al.} 2018,
  \mnras, 481, 414

\bibitem[{{Gonzalez} {et~al.}(2005){Gonzalez}, {Zabludoff}, \&
  {Zaritsky}}]{Gonzales2005}
{Gonzalez}, A.~H., {Zabludoff}, A.~I., \& {Zaritsky}, D. 2005, \apj, 618, 195

\bibitem[{{Gregg} \& {West}(1998)}]{Gregg1998}
{Gregg}, M.~D. \& {West}, M.~J. 1998, \nat, 396, 549

\bibitem[{{Gu} {et~al.}(2018){Gu}, {Conroy}, {Law}, {van Dokkum}, {Yan},
  {Wake}, {Bundy}, {Merritt}, {Abraham}, {Zhang}, {Bershady}, {Bizyaev},
  {Brinkmann}, {Drory}, {Grabowski}, {Masters}, {Pan}, {Parejko}, {Weijmans},
  \& {Zhang}}]{Gu2018}
{Gu}, M., {Conroy}, C., {Law}, D., {et~al.} 2018, \apj, 859, 37

\bibitem[{{Guennou} {et~al.}(2012){Guennou}, {Adami}, {Da Rocha}, {Durret},
  {Ulmer}, {Allam}, {Basa}, {Benoist}, {Biviano}, {Clowe}, {Gavazzi},
  {Halliday}, {Ilbert}, {Johnston}, {Just}, {Kron}, {Kubo}, {Le Brun},
  {Marshall}, {Mazure}, {Murphy}, {Pereira}, {Raba{\c c}a}, {Rostagni},
  {Rudnick}, {Russeil}, {Schrabback}, {Slezak}, {Tucker}, \&
  {Zaritsky}}]{Guennou2012}
{Guennou}, L., {Adami}, C., {Da Rocha}, C., {et~al.} 2012, \aap, 537, A64

\bibitem[{{Hung} \& {Ebeling}(2012)}]{Hung2012}
{Hung}, C.-L. \& {Ebeling}, H. 2012, \mnras, 421, 3229

\bibitem[{{Ilbert} {et~al.}(2006){Ilbert}, {Arnouts}, {McCracken},
  {Bolzonella}, {Bertin}, {Le F{\`e}vre}, {Mellier}, {Zamorani}, {Pell{\`o}},
  {Iovino}, {Tresse}, {Le Brun}, {Bottini}, {Garilli}, {Maccagni}, {Picat},
  {Scaramella}, {Scodeggio}, {Vettolani}, {Zanichelli}, {Adami}, {Bardelli},
  {Cappi}, {Charlot}, {Ciliegi}, {Contini}, {Cucciati}, {Foucaud}, {Franzetti},
  {Gavignaud}, {Guzzo}, {Marano}, {Marinoni}, {Mazure}, {Meneux}, {Merighi},
  {Paltani}, {Pollo}, {Pozzetti}, {Radovich}, {Zucca}, {Bondi}, {Bongiorno},
  {Busarello}, {de La Torre}, {Gregorini}, {Lamareille}, {Mathez}, {Merluzzi},
  {Ripepi}, {Rizzo}, \& {Vergani}}]{Ilbert2006}
{Ilbert}, O., {Arnouts}, S., {McCracken}, H.~J., {et~al.} 2006, \aap, 457, 841

\bibitem[{{Janowiecki} {et~al.}(2010){Janowiecki}, {Mihos}, {Harding},
  {Feldmeier}, {Rudick}, \& {Morrison}}]{Janowiecki2010}
{Janowiecki}, S., {Mihos}, J.~C., {Harding}, P., {et~al.} 2010, \apj, 715, 972

\bibitem[{{Jauzac} {et~al.}(2018){Jauzac}, {Eckert}, {Schaller}, {Schwinn},
  {Massey}, {Bah{\'e}}, {Baugh}, {Barnes}, {Dalla Vecchia}, {Ebeling},
  {Harvey}, {Jullo}, {Kay}, {Kneib}, {Limousin}, {Medezinski}, {Natarajan},
  {Nonino}, {Robertson}, {Tam}, \& {Umetsu}}]{Jauzac2018}
{Jauzac}, M., {Eckert}, D., {Schaller}, M., {et~al.} 2018, \mnras, 481, 2901

\bibitem[{{Jauzac} {et~al.}(2012){Jauzac}, {Jullo}, {Kneib}, {Ebeling},
  {Leauthaud}, {Ma}, {Limousin}, {Massey}, \& {Richard}}]{Jauzac2012}
{Jauzac}, M., {Jullo}, E., {Kneib}, J.-P., {et~al.} 2012, \mnras, 426, 3369

\bibitem[{{Jim{\'e}nez-Teja} \& {Dupke}(2016)}]{Jimenez-Teja2016}
{Jim{\'e}nez-Teja}, Y. \& {Dupke}, R. 2016, \apj, 820, 49

\bibitem[{{Jim{\'e}nez-Teja} {et~al.}(2018){Jim{\'e}nez-Teja}, {Dupke},
  {Ben{\'{\i}}tez}, {Koekemoer}, {Zitrin}, {Umetsu}, {Ziegler}, {Frye}, {Ford},
  {Bouwens}, {Bradley}, {Broadhurst}, {Coe}, {Donahue}, {Graves}, {Grillo},
  {Infante}, {Jouvel}, {Kelson}, {Lahav}, {Lazkoz}, {Lemze}, {Maoz},
  {Medezinski}, {Melchior}, {Meneghetti}, {Mercurio}, {Merten}, {Molino},
  {Moustakas}, {Nonino}, {Ogaz}, {Riess}, {Rosati}, {Sayers}, {Seitz}, \&
  {Zheng}}]{Jimenez-Teja2018}
{Jim{\'e}nez-Teja}, Y., {Dupke}, R., {Ben{\'{\i}}tez}, N., {et~al.} 2018, \apj,
  857, 79

\bibitem[{{Jim{\'e}nez-Teja} {et~al.}(2019){Jim{\'e}nez-Teja}, {Dupke}, {Lopes
  de Oliveira}, {Xavier}, {Coelho}, {Chies-Santos}, {L{\'o}pez-Sanjuan},
  {Alvarez-Candal}, {Costa-Duarte}, {Telles}, {Hernandez-Jimenez},
  {Ben{\'{\i}}tez}, {Alcaniz}, {Cenarro}, {Crist{\'o}bal-Hornillos},
  {Ederoclite}, {Mar{\'{\i}}n-Franch}, {Mendes de Oliveira}, {Moles},
  {Sodr{\'e}}, {Varela}, \& {V{\'a}zquez Rami{\'o}}}]{Jimenez-Teja2019}
{Jim{\'e}nez-Teja}, Y., {Dupke}, R.~A., {Lopes de Oliveira}, R., {et~al.} 2019,
  \aap, 622, A183

\bibitem[{{Karabal} {et~al.}(2017){Karabal}, {Duc}, {Kuntschner}, {Chanial},
  {Cuillandre}, \& {Gwyn}}]{Karabal2017}
{Karabal}, E., {Duc}, P.-A., {Kuntschner}, H., {et~al.} 2017, \aap, 601, A86

\bibitem[{{Kartaltepe} {et~al.}(2008){Kartaltepe}, {Ebeling}, {Ma}, \&
  {Donovan}}]{Kartaltepe2008}
{Kartaltepe}, J.~S., {Ebeling}, H., {Ma}, C.~J., \& {Donovan}, D. 2008, \mnras,
  389, 1240

\bibitem[{{Ko} \& {Jee}(2018)}]{Ko2018}
{Ko}, J. \& {Jee}, M.~J. 2018, \apj, 862, 95

\bibitem[{{Krick} \& {Bernstein}(2007)}]{Krick2007}
{Krick}, J.~E. \& {Bernstein}, R.~A. 2007, \aj, 134, 466

\bibitem[{{Krick} {et~al.}(2006){Krick}, {Bernstein}, \&
  {Pimbblet}}]{Krick2006}
{Krick}, J.~E., {Bernstein}, R.~A., \& {Pimbblet}, K.~A. 2006, \aj, 131, 168

\bibitem[{{Krist} {et~al.}(2011){Krist}, {Hook}, \& {Stoehr}}]{Krist2011}
{Krist}, J.~E., {Hook}, R.~N., \& {Stoehr}, F. 2011, in \procspie, Vol. 8127,
  Optical Modeling and Performance Predictions V, 81270J

\bibitem[{{Krolewski} {et~al.}(2019){Krolewski}, {Ho}, {Chen}, {Chan},
  {Tenneti}, {Bizyaev}, \& {Kraljic}}]{Krolewski2019}
{Krolewski}, A., {Ho}, S., {Chen}, Y.-C., {et~al.} 2019, arXiv e-prints

\bibitem[{{Kuutma} {et~al.}(2017){Kuutma}, {Tamm}, \& {Tempel}}]{Kuutma2017}
{Kuutma}, T., {Tamm}, A., \& {Tempel}, E. 2017, \aap, 600, L6

\bibitem[{{Libeskind} {et~al.}(2018){Libeskind}, {van de Weygaert}, {Cautun},
  {Falck}, {Tempel}, {Abel}, {Alpaslan}, {Arag{\'o}n-Calvo}, {Forero-Romero},
  {Gonzalez}, {Gottl{\"o}ber}, {Hahn}, {Hellwing}, {Hoffman}, {Jones},
  {Kitaura}, {Knebe}, {Manti}, {Neyrinck}, {Nuza}, {Padilla}, {Platen},
  {Ramachandra}, {Robotham}, {Saar}, {Shandarin}, {Steinmetz}, {Stoica},
  {Sousbie}, \& {Yepes}}]{Libeskind2018}
{Libeskind}, N.~I., {van de Weygaert}, R., {Cautun}, M., {et~al.} 2018, \mnras,
  473, 1195

\bibitem[{{Limousin} {et~al.}(2016){Limousin}, {Richard}, {Jullo}, {Jauzac},
  {Ebeling}, {Bonamigo}, {Alavi}, {Cl{\'e}ment}, {Giocoli}, {Kneib}, {Verdugo},
  {Natarajan}, {Siana}, {Atek}, \& {Rexroth}}]{Limousin2016}
{Limousin}, M., {Richard}, J., {Jullo}, E., {et~al.} 2016, \aap, 588, A99

\bibitem[{{Lotz} {et~al.}(2017){Lotz}, {Koekemoer}, {Coe}, {Grogin}, {Capak},
  {Mack}, {Anderson}, {Avila}, {Barker}, {Borncamp}, {Brammer}, {Durbin},
  {Gunning}, {Hilbert}, {Jenkner}, {Khandrika}, {Levay}, {Lucas}, {MacKenty},
  {Ogaz}, {Porterfield}, {Reid}, {Robberto}, {Royle}, {Smith},
  {Storrie-Lombardi}, {Sunnquist}, {Surace}, {Taylor}, {Williams}, {Bullock},
  {Dickinson}, {Finkelstein}, {Natarajan}, {Richard}, {Robertson}, {Tumlinson},
  {Zitrin}, {Flanagan}, {Sembach}, {Soifer}, \& {Mountain}}]{Lotz2017}
{Lotz}, J.~M., {Koekemoer}, A., {Coe}, D., {et~al.} 2017, \apj, 837, 97

\bibitem[{{Lucy}(1974)}]{Lucy1974}
{Lucy}, L.~B. 1974, \aj, 79, 745

\bibitem[{{Mahajan} {et~al.}(2012){Mahajan}, {Raychaudhury}, \&
  {Pimbblet}}]{Mahajan2012}
{Mahajan}, S., {Raychaudhury}, S., \& {Pimbblet}, K.~A. 2012, \mnras, 427, 1252

\bibitem[{{Martinet} {et~al.}(2016){Martinet}, {Clowe}, {Durret}, {Adami},
  {Acebr{\'o}n}, {Hernandez-Garc{\'{\i}}a}, {M{\'a}rquez}, {Guennou}, {Sarron},
  \& {Ulmer}}]{Martinet2016}
{Martinet}, N., {Clowe}, D., {Durret}, F., {et~al.} 2016, \aap, 590, A69

\bibitem[{{Martinet} {et~al.}(2017){Martinet}, {Durret}, {Adami}, \&
  {Rudnick}}]{Martinet2017}
{Martinet}, N., {Durret}, F., {Adami}, C., \& {Rudnick}, G. 2017, \aap, 604,
  A80

\bibitem[{{Martinet} {et~al.}(2015){Martinet}, {Durret}, {Guennou}, {Adami},
  {Biviano}, {Ulmer}, {Clowe}, {Halliday}, {Ilbert}, {M{\'a}rquez}, \&
  {Schirmer}}]{Martinet2015}
{Martinet}, N., {Durret}, F., {Guennou}, L., {et~al.} 2015, \aap, 575, A116

\bibitem[{{Merritt}(1984)}]{Merritt1984}
{Merritt}, D. 1984, \apj, 276, 26

\bibitem[{{Mihos}(2004{\natexlab{a}})}]{Mihos2004b}
{Mihos}, J.~C. 2004{\natexlab{a}}, Clusters of Galaxies: Probes of Cosmological
  Structure and Galaxy Evolution, 277

\bibitem[{{Mihos}(2004{\natexlab{b}})}]{Mihos2004a}
{Mihos}, J.~C. 2004{\natexlab{b}}, in IAU Symposium, Vol. 217, Recycling
  Intergalactic and Interstellar Matter, ed. P.-A. {Duc}, J.~{Braine}, \&
  E.~{Brinks}, 390

\bibitem[{{Mihos} {et~al.}(2005){Mihos}, {Harding}, {Feldmeier}, \&
  {Morrison}}]{Mihos2005}
{Mihos}, J.~C., {Harding}, P., {Feldmeier}, J., \& {Morrison}, H. 2005, \apjl,
  631, L41

\bibitem[{{Mihos} {et~al.}(2017){Mihos}, {Harding}, {Feldmeier}, {Rudick},
  {Janowiecki}, {Morrison}, {Slater}, \& {Watkins}}]{Mihos2017}
{Mihos}, J.~C., {Harding}, P., {Feldmeier}, J.~J., {et~al.} 2017, \apj, 834, 16

\bibitem[{{Montes} \& {Trujillo}(2018)}]{Montes2018}
{Montes}, M. \& {Trujillo}, I. 2018, \mnras, 474, 917

\bibitem[{{Montes} \& {Trujillo}(2019)}]{Montes2019}
{Montes}, M. \& {Trujillo}, I. 2019, \mnras, 482, 2838

\bibitem[{{Moore} {et~al.}(1996){Moore}, {Katz}, {Lake}, {Dressler}, \&
  {Oemler}}]{Moore1996}
{Moore}, B., {Katz}, N., {Lake}, G., {Dressler}, A., \& {Oemler}, A. 1996,
  \nat, 379, 613

\bibitem[{{Moore} {et~al.}(1999){Moore}, {Lake}, {Quinn}, \&
  {Stadel}}]{Moore1999}
{Moore}, B., {Lake}, G., {Quinn}, T., \& {Stadel}, J. 1999, \mnras, 304, 465

\bibitem[{{Morishita} {et~al.}(2017){Morishita}, {Abramson}, {Treu}, {Schmidt},
  {Vulcani}, \& {Wang}}]{Morishita2017}
{Morishita}, T., {Abramson}, L.~E., {Treu}, T., {et~al.} 2017, \apj, 846, 139

\bibitem[{{Murante} {et~al.}(2004){Murante}, {Arnaboldi}, {Gerhard}, {Borgani},
  {Cheng}, {Diaferio}, {Dolag}, {Moscardini}, {Tormen}, {Tornatore}, \&
  {Tozzi}}]{Murante2004}
{Murante}, G., {Arnaboldi}, M., {Gerhard}, O., {et~al.} 2004, \apjl, 607, L83

\bibitem[{{Murante} {et~al.}(2007){Murante}, {Giovalli}, {Gerhard},
  {Arnaboldi}, {Borgani}, \& {Dolag}}]{Murante2007}
{Murante}, G., {Giovalli}, M., {Gerhard}, O., {et~al.} 2007, \mnras, 377, 2

\bibitem[{{Napolitano} {et~al.}(2003){Napolitano}, {Pannella}, {Arnaboldi},
  {Gerhard}, {Aguerri}, {Freeman}, {Capaccioli}, {Ghigna}, {Governato},
  {Quinn}, \& {Stadel}}]{Napolitano2003}
{Napolitano}, N.~R., {Pannella}, M., {Arnaboldi}, M., {et~al.} 2003, \apj, 594,
  172

\bibitem[{{Planck Collaboration} {et~al.}(2013){Planck Collaboration}, {Ade},
  {Aghanim}, {Arnaud}, {Ashdown}, {Atrio-Barandela}, {Aumont}, {Baccigalupi},
  {Balbi}, {Banday}, \& et~al.}]{Planck2013}
{Planck Collaboration}, {Ade}, P.~A.~R., {Aghanim}, N., {et~al.} 2013, \aap,
  550, A134

\bibitem[{{Ricci} {et~al.}(2018){Ricci}, {Benoist}, {Maurogordato}, {Adami},
  {Chiappetti}, {Gastaldello}, {Guglielmo}, {Poggianti}, {Sereno}, {Adam},
  {Arnouts}, {Cappi}, {Koulouridis}, {Pacaud}, {Pierre}, \&
  {Ramos-Ceja}}]{Ricci2018}
{Ricci}, M., {Benoist}, C., {Maurogordato}, S., {et~al.} 2018, \aap, 620, A13

\bibitem[{Richardson(1972)}]{Richardson1972}
Richardson, W.~H. 1972, J. Opt. Soc. Am., 62, 55

\bibitem[{{Rong} {et~al.}(2016){Rong}, {Liu}, \& {Zhang}}]{Rong2016}
{Rong}, Y., {Liu}, Y., \& {Zhang}, S.-N. 2016, \mnras, 455, 2267

\bibitem[{{Rudick} {et~al.}(2009){Rudick}, {Mihos}, {Frey}, \&
  {McBride}}]{Rudick2009}
{Rudick}, C.~S., {Mihos}, J.~C., {Frey}, L.~H., \& {McBride}, C.~K. 2009, \apj,
  699, 1518

\bibitem[{{Rudick} {et~al.}(2006){Rudick}, {Mihos}, \& {McBride}}]{Rudick2006}
{Rudick}, C.~S., {Mihos}, J.~C., \& {McBride}, C. 2006, \apj, 648, 936

\bibitem[{{Rudick} {et~al.}(2011){Rudick}, {Mihos}, \& {McBride}}]{Rudick2011}
{Rudick}, C.~S., {Mihos}, J.~C., \& {McBride}, C.~K. 2011, \apj, 732, 48

\bibitem[{{Rudnick} {et~al.}(2009){Rudnick}, {von der Linden}, {Pell{\'o}},
  {Arag{\'o}n-Salamanca}, {Marchesini}, {Clowe}, {De Lucia}, {Halliday},
  {Jablonka}, {Milvang-Jensen}, {Poggianti}, {Saglia}, {Simard}, {White}, \&
  {Zaritsky}}]{Rudnick2009}
{Rudnick}, G., {von der Linden}, A., {Pell{\'o}}, R., {et~al.} 2009, \apj, 700,
  1559

\bibitem[{{Sandin}(2014)}]{Sandin2014}
{Sandin}, C. 2014, \aap, 567, A97

\bibitem[{{Sandin}(2015)}]{Sandin2015}
{Sandin}, C. 2015, \aap, 577, A106

\bibitem[{{Sarron} {et~al.}(2019){Sarron}, {Adami}, {Durret}, \&
  {Laigle}}]{Sarron2019}
{Sarron}, F., {Adami}, C., {Durret}, F., \& {Laigle}, C. 2019, arXiv e-prints

\bibitem[{{Sarron} {et~al.}(2018){Sarron}, {Martinet}, {Durret}, \&
  {Adami}}]{Sarron2018}
{Sarron}, F., {Martinet}, N., {Durret}, F., \& {Adami}, C. 2018, \aap, 613, A67

\bibitem[{{Schechter}(1976)}]{Schechter1976}
{Schechter}, P. 1976, \apj, 203, 297

\bibitem[{{Schlafly} \& {Finkbeiner}(2011)}]{Schlafly2011}
{Schlafly}, E.~F. \& {Finkbeiner}, D.~P. 2011, \apj, 737, 103

\bibitem[{Shensa(1992)}]{Shensa1992}
Shensa, M.~J. 1992, {IEEE} Trans. Signal Processing, 40, 2464

\bibitem[{{Simard} {et~al.}(2009){Simard}, {Clowe}, {Desai}, {Dalcanton}, {von
  der Linden}, {Poggianti}, {White}, {Arag{\'o}n-Salamanca}, {De Lucia},
  {Halliday}, {Jablonka}, {Milvang-Jensen}, {Saglia}, {Pell{\'o}}, {Rudnick},
  \& {Zaritsky}}]{Simard2009}
{Simard}, L., {Clowe}, D., {Desai}, V., {et~al.} 2009, \aap, 508, 1141

\bibitem[{{Skelton} {et~al.}(2014){Skelton}, {Whitaker}, {Momcheva}, {Brammer},
  {van Dokkum}, {Labb{\'e}}, {Franx}, {van der Wel}, {Bezanson}, {Da Cunha},
  {Fumagalli}, {F{\"o}rster Schreiber}, {Kriek}, {Leja}, {Lundgren}, {Magee},
  {Marchesini}, {Maseda}, {Nelson}, {Oesch}, {Pacifici}, {Patel}, {Price},
  {Rix}, {Tal}, {Wake}, \& {Wuyts}}]{Skelton2014}
{Skelton}, R.~E., {Whitaker}, K.~E., {Momcheva}, I.~G., {et~al.} 2014, \apjs,
  214, 24

\bibitem[{{Smail} {et~al.}(1998){Smail}, {Edge}, {Ellis}, \&
  {Blandford}}]{Smail1998}
{Smail}, I., {Edge}, A.~C., {Ellis}, R.~S., \& {Blandford}, R.~D. 1998, \mnras,
  293, 124

\bibitem[{{Sommer-Larsen}(2006)}]{Sommer-larsen2006}
{Sommer-Larsen}, J. 2006, \mnras, 369, 958

\bibitem[{{Starck} {et~al.}(1998){Starck}, {Murtagh}, \&
  {Bijaoui}}]{Starck1998}
{Starck}, J.-L., {Murtagh}, F.~D., \& {Bijaoui}, A. 1998, {Image Processing and
  Data Analysis}, 297

\bibitem[{{Stoica} {et~al.}(2005){Stoica}, {Mart{\'{\i}}nez}, {Mateu}, \&
  {Saar}}]{Stoica2005}
{Stoica}, R.~S., {Mart{\'{\i}}nez}, V.~J., {Mateu}, J., \& {Saar}, E. 2005,
  \aap, 434, 423

\bibitem[{{Tang} {et~al.}(2018){Tang}, {Lin}, {Cui}, {Kang}, {Wang}, {Contini},
  \& {Yu}}]{Tang2018}
{Tang}, L., {Lin}, W., {Cui}, W., {et~al.} 2018, \apj, 859, 85

\bibitem[{{Taylor}(2005)}]{Taylor2005}
{Taylor}, M.~B. 2005, in Astronomical Society of the Pacific Conference Series,
  Vol. 347, Astronomical Data Analysis Software and Systems XIV, ed.
  P.~{Shopbell}, M.~{Britton}, \& R.~{Ebert}, 29

\bibitem[{{Tempel} {et~al.}(2015){Tempel}, {Guo}, {Kipper}, \&
  {Libeskind}}]{Tempel2015}
{Tempel}, E., {Guo}, Q., {Kipper}, R., \& {Libeskind}, N.~I. 2015, \mnras, 450,
  2727

\bibitem[{{Tempel} \& {Libeskind}(2013)}]{Tempel2013a}
{Tempel}, E. \& {Libeskind}, N.~I. 2013, \apjl, 775, L42

\bibitem[{{Tempel} {et~al.}(2016){Tempel}, {Stoica}, {Kipper}, \&
  {Saar}}]{Tempel2016}
{Tempel}, E., {Stoica}, R.~S., {Kipper}, R., \& {Saar}, E. 2016, Astronomy and
  Computing, 16, 17

\bibitem[{{Tempel} {et~al.}(2014){Tempel}, {Stoica}, {Mart{\'{\i}}nez},
  {Liivam{\"a}gi}, {Castellan}, \& {Saar}}]{Tempel2014}
{Tempel}, E., {Stoica}, R.~S., {Mart{\'{\i}}nez}, V.~J., {et~al.} 2014, \mnras,
  438, 3465

\bibitem[{{Tempel} {et~al.}(2013){Tempel}, {Stoica}, \& {Saar}}]{Tempel2013b}
{Tempel}, E., {Stoica}, R.~S., \& {Saar}, E. 2013, \mnras, 428, 1827

\bibitem[{{Trentham} \& {Mobasher}(1998)}]{Trentham1998}
{Trentham}, N. \& {Mobasher}, B. 1998, \mnras, 293, 53

\bibitem[{{Trujillo} \& {Fliri}(2016)}]{Trujillo2016}
{Trujillo}, I. \& {Fliri}, J. 2016, \apj, 823, 123

\bibitem[{{Vacca} {et~al.}(2018){Vacca}, {Murgia}, {Govoni}, {Loi}, {Vazza},
  {Finoguenov}, {Carretti}, {Feretti}, {Giovannini}, {Concu}, {Melis},
  {Gheller}, {Paladino}, {Poppi}, {Valente}, {Bernardi}, {Boschin}, {Brienza},
  {Clarke}, {Colafrancesco}, {En{\ss}lin}, {Ferrari}, {de Gasperin},
  {Gastaldello}, {Girardi}, {Gregorini}, {Johnston-Hollitt}, {Junklewitz},
  {Orr{\`u}}, {Parma}, {Perley}, \& {Taylor}}]{Vacca2018}
{Vacca}, V., {Murgia}, M., {Govoni}, F., {et~al.} 2018, \mnras, 479, 776

\bibitem[{{Vilchez-Gomez} {et~al.}(1994){Vilchez-Gomez}, {Pello}, \&
  {Sanahuja}}]{Vilchez-gomez1994}
{Vilchez-Gomez}, R., {Pello}, R., \& {Sanahuja}, B. 1994, \aap, 283, 37

\bibitem[{{Vulcani} {et~al.}(2011){Vulcani}, {Poggianti},
  {Arag{\'o}n-Salamanca}, {Fasano}, {Rudnick}, {Valentinuzzi}, {Dressler},
  {Bettoni}, {Cava}, {D'Onofrio}, {Fritz}, {Moretti}, {Omizzolo}, \&
  {Varela}}]{Vulcani2011}
{Vulcani}, B., {Poggianti}, B.~M., {Arag{\'o}n-Salamanca}, A., {et~al.} 2011,
  \mnras, 412, 246

\bibitem[{{Wang} {et~al.}(2018){Wang}, {Guo}, {Kang}, \&
  {Libeskind}}]{Wang2018}
{Wang}, P., {Guo}, Q., {Kang}, X., \& {Libeskind}, N.~I. 2018, \apj, 866, 138

\bibitem[{{Willman} {et~al.}(2004){Willman}, {Governato}, {Wadsley}, \&
  {Quinn}}]{Willman2004}
{Willman}, B., {Governato}, F., {Wadsley}, J., \& {Quinn}, T. 2004, \mnras,
  355, 159

\bibitem[{{Zeldovich} {et~al.}(1982){Zeldovich}, {Einasto}, \&
  {Shandarin}}]{Zeldovich1982}
{Zeldovich}, I.~B., {Einasto}, J., \& {Shandarin}, S.~F. 1982, \nat, 300, 407

\bibitem[{{Zenteno} {et~al.}(2016){Zenteno}, {Mohr}, {Desai}, {Stalder},
  {Saro}, {Dietrich}, {Bayliss}, {Bocquet}, {Chiu}, {Gonzalez}, {Gangkofner},
  {Gupta}, {Hlavacek-Larrondo}, {McDonald}, {Reichardt}, \&
  {Rest}}]{Zenteno2016}
{Zenteno}, A., {Mohr}, J.~J., {Desai}, S., {et~al.} 2016, \mnras, 462, 830

\bibitem[{{Zhang} {et~al.}(2015){Zhang}, {Yang}, {Wang}, {Wang}, {Luo}, {Mo},
  \& {van den Bosch}}]{Zhang2015}
{Zhang}, Y., {Yang}, X., {Wang}, H., {et~al.} 2015, \apj, 798, 17

\bibitem[{{Zhang} {et~al.}(2019){Zhang}, {Yanny}, {Palmese}, {Gruen}, {To},
  {Rykoff}, {Leung}, {Collins}, {Hilton}, {Abbott}, {Annis}, {Avila}, {Bertin},
  {Brooks}, {Burke}, {Carnero Rosell}, {Carrasco Kind}, {Carretero}, {Cunha},
  {D'Andrea}, {da Costa}, {De Vicente}, {Desai}, {Diehl}, {Dietrich},
  {Doel}, {Drlica-Wagner}, {Eifler}, {Evrard}, {Flaugher}, {Fosalba},
  {Frieman}, {Garc{\'{\i}}a-Bellido}, {Gaztanaga}, {Gerdes}, {Gruendl},
  {Gschwend}, {Gutierrez}, {Hartley}, {Hollowood}, {Honscheid}, {Hoyle},
  {James}, {Jeltema}, {Kuehn}, {Kuropatkin}, {Li}, {Lima}, {Maia}, {March},
  {Marshall}, {Melchior}, {Menanteau}, {Miller}, {Miquel}, {Mohr}, {Ogando},
  {Plazas}, {Romer}, {Sanchez}, {Scarpine}, {Schubnell}, {Serrano},
  {Sevilla-Noarbe}, {Smith}, {Soares-Santos}, {Sobreira}, {Suchyta}, {Swanson},
  {Tarle}, {Thomas}, {Wester}, \& {DES Collaboration}}]{Zhang2019}
{Zhang}, Y., {Yanny}, B., {Palmese}, A., {et~al.} 2019, \apj, 874, 165

\bibitem[{{Zwicky}(1951)}]{Zwicky1951}
{Zwicky}, F. 1951, \pasp, 63, 61

\end{thebibliography}

\end{document}